\documentclass[cmp,final]{svjour}

\journalname{Communications in Mathematical Physics}

\makeatletter

\@addtoreset{equation}{section}
\makeatother

\def\ds{\displaystyle}

\def\Od{\Omega_\delta}

\def\e{\varepsilon}
\def\ov{\overline}

\def\ti{\tilde}
\def\d{\partial}
\def\la{\langle}
\def\ra{\rangle}
\def\const{\,\mbox{const}\,}

\def\bx{\mbox{\boldmath$\xi$}}
\def\bs{\mbox{\boldmath$\sigma$}}
\def\un{u^{\nu}}
\def\um{u^{\mu}}
\def\xm{\xi^{\mu}}
\def\vm{v^{\mu}}
\def\vn{v^{\nu}}
\def\wn{w^\nu}
\def\wm{w^\mu}
\def\ln{\lambda^{\nu}}
\def\lm{\lambda^{\mu}}
\def\tn{t^{\nu}}
\def\tm{t^{\mu}}
\def\tv{\ov t}
\def\vv{\ov v}
\def\uv{\ov u}
\def\wv{\ov w}
\def\lv{\ov \lambda}
\def\zj{\zeta_j}
\def\zk{\zeta_k}
\def\s{\sigma}
\def\l{\lambda}
\def\hN{\chi_{N,h}}
\def\hhN{\hat\chi_{N,h}}
\def\no{\noindent}
\def\de{\delta}

\def\xmj{\xi^{\mu}_j}
\def\xmk{\xi^{\mu}_k}

\def\P{\hbox{Prob}}
\def\F{{\cal F}_0}
\def\a{\alpha}

\begin{document}
\title{{On the Critical Capacity of the Hopfield Model}}
\titlerunning{On the Critical Capacity of the Hopfield Model}

\author{Jianfeng Feng\inst{1}\and Mariya Shcherbina\inst{2}\and
 Brunello Tirozzi\inst{3}}

\institute{Laboratory of Neurocomputation, The Babraham
Institute, Cambridge, CB2 4AT, UK,  \email{jf218@hermes.cam.ac.uk}
\and Institute for Low
Temperature Physics,Ukr. Ac. Sci., 47 Lenin ave., Kharkov,
Ukraine, \email{shcherbi@ilt.kharkov.ua}\and Department of Physics of
 Rome University "La Sapienza", 5, p-za A.Moro, Rome, Italy,
\email{tirozzi@krishna.phys.uniroma1.it}}

\date{Received: 1 December 1999/ Accepted: 21 July 2000}
\communicated{Ya.G.Sinai}

\maketitle

\begin{abstract}
We estimate the critical capacity of the zero-temperature Hopfield
model by using a novel and rigorous method. The probability of
having a stable fixed point is one when $\alpha\le 0.113$ for a
large number of neurons. This result is an advance on all rigorous
results in the literature and the relationship  between the
capacity $\alpha$ and retrieval errors  obtained here for small
$\alpha$ coincides with replica calculation results.
\end{abstract}

\section{Introduction and Main Results}

The Hopfield model is one of the most important
models in the theory of spin glasses and neural networks \cite{Hop,MPV}.
It has been intensively investigated in the past few
years (see e.g. book \cite{MPV} and references therein).
One of the main problems is the critical
capacity which has been studied by means of the replica trick \cite{A,Amit}. 
Here the value $\a_c=0.138...$ (coinciding also with
numerical experiments) was found. But this result is nonrigorous from
the  mathematical point of view.
There are few   rigorous approaches in the literature to
estimate the critical capacity of the Hopfield model
\cite{N,Lou,Tal}.
 Here we introduce a novel approach based upon
analysis of the Fourier transform of the joint distribution of the
effective fields. It enables us to obtain a new bound for the critical
capacity and also allows us to prove rigorously, for small
$\alpha$, the results obtained in terms of the extreme value
theory \cite{FeT}.

Consider the sequential dynamics of the Hopfield model in the form

\begin{equation}
\sigma_k(t+1)=\hbox{sign}\{\sum_{j=1,j\not=k}^N \ti J_{kj}\sigma_j(t)\},
\label{dyn}\end{equation}
where, as usual,
\begin{equation}
\ti J_{jk}={1\over N}\sum_{\mu=1}^{p+1}\ti\xmj\ti\xmk, \quad
{p\over N}\to \alpha,\quad as \quad N\to \infty,
\label{tiJ}\end{equation} and $\ti\xm_k$ $(j=1,\dots,N)$,
$(\mu=1,\dots,p+1)$ are i.i.d. random variables assuming values
$\pm 1$ with probability ${1\over 2}$.  This dynamical system is
determined by the energy function
\begin{equation}
{\cal H}(\bs)=-{1\over 2}\sum_{j\not=k}^N\ti J_{jk}\s_j\s_k,
\label{H(s)}\end{equation} where we denote
$\bs\equiv(\s_1,\dots,\s_N)$. It is easily seen that the function
${\cal H}(\bs)$ does not increase in the
 process of evolution. Thus, the dynamics
of the model depends on the "energy landscape" of the function
${\cal H}(\bs)$ and the local minima of the function are the fixed
points of dynamics  (\ref{dyn}). Newman \cite{N} was the first,
who proved, that for $\alpha\le 0.056.$, an
 "energy barrier" exists with probability 1 around every point
 $\bs^\mu=\bx^\mu\equiv
(\ti\xm_1,\dots,\ti\xm_N)$, i.e.
 there exist some positive numbers $\delta$ and $\e$, such that for any
$\bs$, belonging to 
$$ \Od^\mu\equiv\{\bs:||\bs-\bx^\mu||^2=
2[\delta N]\}, $$ 
the following inequality holds:
$$ {\cal H}(\bs)-{\cal H}(\bx^\mu)\ge \e N $$ 
(here and below the norm
$||...||$ corresponds to the usual scalar product $(...,...)$ in
${\bf R}^N$). In other words, it means that
\begin{equation}
\min_{\bs\in\Od^\mu}{\cal H}(\bs)-{\cal H}(\bx^\mu)\ge \e^2 N.
\label{min}\end{equation}
This result was improved by Loukianova \cite{Lou}, who proved the existence of the
"energy barriers" for $\alpha\le 0.071$ and then by Talagrand \cite{Tal}.
One can show, that if such a "barrier" exists, then inside
each open ball
$$
B^\mu_\delta\equiv\{\bs:||\bs-\bx^\mu||^2<2[\delta N]\}
$$
there exists a point of local minimum of the function ${\cal H}(\bs)$, which,
as it was mentioned above, is the fixed point of dynamics (\ref{dyn}).

Thus, it is clear that the point $\bs^*$ in which
${\cal H}(\bs^*)=\min_{\bs\in\Od^\mu}{\cal H}(\bs)$ plays an important
role in dynamics (\ref{dyn}).
We shall  study the probability of the event, that the
point $\bs^{(1,\delta)}\in\Od^1$ with
\begin{equation}
\sigma_k^{(1,\delta)}=-\ti\xi^1_k,\,\,\, (k=1,\dots,[\delta
N]),\quad \sigma_k^{(1,\delta)}=\ti\xi^1_k,\,\,\, (k=1+[\delta
N],\dots,N) \label{S^*}\end{equation} is a local minimum of the
function ${\cal H}(\bs)$ on $\Od^1$. This means that ${\cal
H}(\bs^{(1,\delta)})$ must be less than the value of ${\cal
H}(\bs)$ for any $\bs\in\Od^1$ which is the ``nearest neighbor''
of $\bs^{(1,\delta)}$ in $\Od^1$. It is easy to see that,  it is
so
 if and only if for any $k=1,\dots,[\delta N]$ and $j=[\delta N]+1,\dots,N$,
\begin{equation}
-2\ti J_{kj}\sigma^{(1,\delta)}_j\sigma^{(1,\delta)}_k+
\sigma^{(1,\delta)}_k\sum_{i=1,i\not= k}^N\ti
J_{ki}\sigma^{(1,\delta)}_i+
\sigma^{(1,\delta)}_j\sum_{i=1,i\not=j }^N\ti
J_{ji}\sigma^{(1,\delta)}_i \ge 0.
\label{cond0}\end{equation}
It is useful to introduce  at this point the definition of
``effective fields''.
\begin{definition}
The effective fields generated by the configuration $\bs$ on the
neuron $k$ is
$$
z_k\equiv\sigma_k\sum_{i=1,i\not= k}^N\ti
J_{ki}\sigma_i.
$$
\end{definition}

Our approach is based on the analysis of the joint probability
distribution of the variables $z_k$ $(k=1,\dots,N)$.

Since with probability
larger than $1-e^{-N\const\ti\e^2}$ all matrix elements $\ti
J_{kj}$ satisfy the inequality
\begin{equation}
|\ti J_{kj}|\le{\ti\e\over 2}\qquad (k,j=1,...,N),
\label{ineqJ}\end{equation}
one can derive from (\ref{cond0}) that, if we denote by $\ti x_k^0$ the effective fields,
generated by the configuration $\bs^{(1,\delta)}$
\begin{equation}
\ti x_k^0=\sigma^{(1,\delta)}_k\sum_{i=1,i\not=k}^N\ti J_{ki}
\sigma^{(1,\delta)}_i, \label{x*}\end{equation}
  the necessary
condition for $\bs^{(1,\delta)}$ to be a local minimum point is
\begin{equation}
\min_{k=1,\dots,[\delta N]}\ti x_k^0+
\min_{j=[\delta N]+1,\dots,N}\ti x_j^0\ge -\ti\e,
\label{condmin}\end{equation}
and the sufficient condition has the same form with $+\ti\e$ in the r.h.s.
 Thus, if we consider the events
\begin{equation}
{\cal A}_k^0( q)=\{\ti x_k^0\ge  q\},
\label{calA}\end{equation}
then the event ${\cal M}$ that $\bs^{(1,\delta)}$ is a local minimum point
satisfies the relations
\begin{equation}\begin{array}{c}\ds{
\cup_{q+ q'\ge \ti\e}(\cap_{k=1}^{[\delta N]}
{\cal A}_k^0( q)\cap_{k=[\delta N]+1}^N{\cal A}_k^0( q')
\subset{\cal M}}\\
\ds{
\subset\cup_{ q+ q'\ge -\ti\e}(\cap_{k=1}^{[\delta N]}
{\cal A}_k^0( q)\cap_{k=[\delta N]+1}^N{\cal A}_k^0( q')). }
\end{array}\label{condmin1}\end{equation}
So we should study the behaviour of
\begin{equation}
P_N( q, q')\equiv\P\{\cap_{k=1}^{[\delta N]} {\cal A}_k^0(
q)\cap_{k=[\delta N]+1}^N{\cal A}_k^0( q')\}.
\label{P(q,q')}\end{equation}
Observe  that, in particular,
$P_N(0,0)$ is the probability to have a fixed point of dynamics
(\ref{dyn}) at the point $\bs^{(1,\delta)}$. Now let us introduce the
new notation:
\begin{equation}
\xm_k\equiv\sigma^{(1,\delta)}_k\ti\xi^{\mu+1}_k,\quad
(\mu=1,...p,\,k=1,...N) .
\label{not}\end{equation}
Then $\xm_k$ $(k=1,\dots,N)$, $(\mu=1,\dots,p)$ are also i.i.d.
random variables assuming the values $\pm 1$ with probability
${1\over 2}$. Denote
\begin{equation}
\ti x_k={1\over N}\sum_{\mu=1}^p\sum_{j=1}^N\xmk\xmj=\ti x_k^0+\alpha_N\pm
(1-2\delta_N),\quad \alpha_N={p+1\over N},\quad \delta_N={[\delta N]\over N}.
\label{x_k}\end{equation}
Here $\alpha_N$ appears because we include in the summation the term
with $j=k$, the term $\pm(1-2\delta_N)$ is due to the term
$N^{-1}(\bx^1,\bs^{(1,\delta)})$, and the sign here depends on $k$:
it is plus for $k=1,\dots,[\delta N]$ and minus for
$k=[\delta N]+1,\dots N$.

To simplify formulae we introduce also
\begin{equation} \begin{array}{c}
 a_{1}\equiv\alpha_N+1-2\delta_N+q\to a_1^*,\quad  a_1^*\equiv
\alpha+1-2\delta+q,\\
\quad a_2\equiv\alpha_N-1+2\delta_N+q'\to a_2^*, \quad
a_2^*\equiv \alpha-1+2\delta+q',
\end{array}\label{a}\end{equation}
which yield
\begin{equation}
 P_N( q, q')\equiv\la\prod_{k=1}^{[\delta N]}\theta(\ti x_k-a_1)
\prod_{k=1+[\delta N]}^N\theta(\ti x_k-a_2)\ra.
\label{tiP_1}\end{equation}
Here and below the symbol
$\la\dots\ra$ denotes averaging with respect to all $\{\xm_k\}$
($k=1,\cdots,N, \mu=1,\cdots,p+1$).

\bigskip

In order to formulate the main results of the paper we need some
other definitions.

Consider the function ${\cal F}_0(U,V;\a,\delta,q,q')$ of the form
\begin{equation}\begin{array}{c}\ds{
{\cal F}_0(U,V;\a,\delta,q,q')\equiv\delta\log
H({a_1^*\over U}-V)+(1-\delta)\log H({a_2^*\over U}-V)}\\
\ds{
-UV+{1\over 2}V^2+\alpha\log U, }
\end{array}\label{F_0}\end{equation}
where
\begin{equation}
H(x)\equiv{1\over\sqrt{2\pi}}\int_{x}^\infty e^{-t^2/2}dt.
\label{H(x)}\end{equation}

Define also
\begin{equation}
A(x)\equiv -{d\over dx}\log H(x)={e^{-x^2/2}\over\sqrt{2\pi}H(x)},
\label{A(x)}\end{equation}
\begin{equation}\begin{array}{c}\ds{
A_{1,2}(U,V)\equiv{1\over U}A({a_{1,2}^*\over U}-V),}\\
\ds{
D(U,V)\equiv{1\over 2}-\delta A_1(U,V)-(1-\delta)A_2(U,V)}\\
\ds{
-{1\over 2}\delta
(1-\delta)(A_1(U,V)-A_2(U,V))^2},
\end{array}\label{D(U)}\end{equation}
and
\begin{equation}
{\cal F}_0^D(U,V;\a,\delta,q,q')\equiv \left\{\begin{array}{l}
{\cal F}_0(U,V;\a,\delta,q,q'),\qquad \mbox{ if }D(U,V)\ge 0\\
\ds{
{1\over 1-2D(U,V)}[\delta\log H({a_1^*\over U}-V)}\\
\ds{
+(1-\delta)\log H({a_2^*\over U}-V)]-UV+{V^2\over 2}+\alpha\log U,}\\
\ds{\hskip 3.4cm \mbox{ if } D(U,V)< 0. }
\end{array}\right.
\label{F^D}\end{equation}
\begin{theorem}\label{thm:1}
\begin{equation} \begin{array}{c}\ds{
\limsup_{N\to \infty}{1\over
N}\log\la\prod_{k=1}^{[\delta N]}\theta(\ti x_k-a_1)
\prod_{k=1+[\delta N]}^N\theta(\ti x_k-a_2)\ra}\\
\ds{
\le\max_{U>0}\min_{V}{\cal F}_0^D(U,V;\a,\delta,q,q')-{\alpha\over 2}\log\alpha+
{\alpha\over 2}. }
\end{array}\label{mest}\end{equation}
\end{theorem}

\begin{remark}
 Note that in all interesting cases
 (see Theorems \ref{thm:2} and \ref{thm:3} below)
$$
\max_{U>0}\min_V{\cal F}_0^D(U,V;\a,\delta,q,q')=\max_{U>0}\min_V
{\cal F}_0(U,V;\a,\delta,q,q')
$$
and one can substitute ${\cal F}_0^D$ by ${\cal F}_0$ in the
r.h.s. of (\ref{mest}).
\end{remark}
\begin{remark}
 The proof of Theorem \ref{thm:1} can be generalized almost literally to  the
 case ( cf. (\ref{tiP_1}))
\begin{equation}\begin{array}{c}\ds{
P_{N,[\delta_1 N]}( q, q')\equiv\la\prod_{k=1+[\delta_1 N]}^{[\delta N]}\theta(\ti x_k-a_1)
\prod_{k=1+[\delta N]}^N\theta(\ti x_k-a_2)\ra}.
\end{array}\label{tiP_2}\end{equation}
We obtain
\begin{equation} \begin{array}{c}\ds{
\limsup_{N\to \infty}{1\over
N}\log P_{N,[\delta_1 N]}( q, q')\le }\\
\ds{
\max_{U>0}\min_V{\cal F}_1^D(U,V;\a,\delta,\delta_1,q,q')-
{\alpha\over 2}\log\alpha+{\alpha\over 2},}
\end{array}\label{mest2}\end{equation}
with (cf. (\ref{F_0})-(\ref{F^D}))
\begin{equation}
{\cal F}_1^D(U,V;\a,\delta,\de_1,q,q')\equiv \left\{\begin{array}{l} {\cal
F}_1(U,V;\a,\delta,\delta_1,q,q'), \qquad \mbox{ if } D^1(U,V)\ge
0;\\ \ds{ {1\over 1-2D^1(U,V)}[(\delta-\delta_1)\log H({a_1^*\over
U}-V)}\\
\ds{
+(1-\delta)\log H({a_2^*\over U}-V)] -UV+{1\over 2}V^2+\alpha\log U,}\\
\ds{\hskip 3.8cm \mbox { if } D^1(U,V)\le 0; }
\end{array}\right.
\label{F^D_1}\end{equation}
where
\begin{equation} \begin{array}{c}\ds{
{\cal F}_1(U,V;\a,\delta,\delta_1,q,q')\equiv
(\delta-\delta_1)\log
H({a_1^*\over U}-V)}\\
\ds{
+(1-\delta)\log H({a_2^*\over U}-V)
-UV+{1\over 2}V^2+\alpha\log U, }
\end{array}\label{F_delta}\end{equation}
and
\begin{equation}\begin{array}{c}\ds{
D^1(U,V)\equiv(1-\delta_1)^{-1}[{1\over 2}-(\delta-\delta_1) A_1(U,V)
-(1-\delta)A_2(U,V)}\\
\ds{
-{1\over 2}(\delta-\delta_1)
(1-\delta)(A_1(U,V)-A_2(U,V))^2)] }
\end{array}\label{D_1(U)}\end{equation}
with $A_{1,2}(U,V)$  defined  in (\ref{D(U)}).
\end{remark}
\medskip

\begin{theorem} \label{thm:2}
If $\alpha$ is small enough, $\delta<<\alpha^3\log\alpha^{-1}$ and
$q=q'=0$, then
\begin{equation}\begin{array}{c}\ds{
\limsup_{N\to \infty}{1\over
N}\log\la\prod_{k=1}^{[\delta N]}\theta(\ti x_k-a_1)
\prod_{k=1+[\delta N]}^N\theta(\ti x_k-a_2)\ra\le
\delta\log
H({1-2\delta\over\sqrt\alpha})}\\
\ds{
+(1-\delta)\log H(-{1-2\delta\over
\sqrt\alpha})+O(e^{-1/\alpha})+o(\delta\log\alpha^{-1}).}
\end{array}\label{esta}\end{equation}
Thus, $P_N^*(\delta, \alpha)$ - the probability to have a fixed
point of the dynamics of the Hopfield model at the  distance $\delta$
from the first pattern has an upper bound of the form:
$$\begin{array}{c}\ds{ P_N^*(\delta,\alpha)\le
\exp\{N[-\delta\log\delta-(1-\delta) \log(1-\delta)+\delta\log
H({1-2\delta\over\sqrt\alpha})}\\
 \ds{
+(1-\delta)\log H(-{1-2\delta\over \sqrt\alpha})
+O(e^{-1/\alpha})+o(\delta\log\alpha^{-1})+o(1)]\}.}
\end{array}$$
 \end{theorem}

\begin{remark}
 It follows from Theorem \ref{thm:2}, that $\delta_c(\alpha)$- the
minimal $\delta$ for which $P_N^*(\delta,\alpha)$ does not decay
exponentially in $N$, as $N\to\infty$, has the asymptotic behaviour
$$ \delta_c(\alpha)\sim{\sqrt\alpha\over\sqrt{2\pi}}
e^{-1/2\alpha}. $$ This result coincides with the formula found by
Amit at al. with
 replica calculations \cite{Amit} and the one, obtained by J.Feng and
B.Tirozzi in \cite{FeT}, using the extreme value theory.
\end{remark}
\medskip

\begin{theorem} \label{thm:3}
Denote by ${\cal A}$ the event that there exist some $\delta,\e>0$
and some point $\bs^0\in B^1_\delta$, such that
$\min_{\bs\in\Od^1}{\cal H}(\bs)- {\cal H}(\bs^0)>\e^2N$.

 Then 
if  for some $\a$ and $\de$ 
\begin{equation}
\max_{0\le q}\max_U\min_V\{{\cal F}_0^D(U,V;\a,\delta, q,
 -q)\}-{\a\over 2}\log\a+{\a\over 2}+C^*(\delta)<0,
\label{T:3.cond}\end{equation}
then there exists some $C(\alpha)>0$ such that
\begin{equation}
\P\{\ov{\cal A}\}\le e^{-NC(\alpha)}.
\label{thm3}\end{equation}
Here and below 
\begin{equation}
C^*(\delta)\equiv -\de\log\de-(1-\de)\log(1-\de).
\label{C^*}\end{equation}
Numerical calculations show that condition (\ref{T:3.cond}) is fulfilled
for any $\a\le\a_c= 0.113...$.
 \end{theorem}
 The paper is organized as follows. In Sec.2 we prove Theorems 1, 2 and 3.
In the process of the proof we shall need some auxiliary facts which
we formulate there as Lemmas 1-4 and Propositions 1-4. Section 3 is
devoted to the  proof of the auxiliary results.

\section{Proof of   Main Results}

{\it Proof of Theorem \ref{thm:1}.}
To make the idea of the proof more understandable we first
carry out   all  computations when $\{\xi^\mu_j\}$ are Gaussian
 random variables.
Since this part has no connection with the rigorous
proof of Theorem \ref{thm:1}, we just sketch the proof, without going into
details.

To find $P^g_N$ which corresponds to $ P_N$ (see (\ref{tiP_1})) in the
Gaussian case, we study the Fourier transform of the joint probability
distribution of the variables $\ti x_k$,
\begin{equation}\begin{array}{c}\ds{
F(\zeta_1,...,\zeta_N)\equiv(2\pi)^{-N/2}\la\exp\{i\sum_{k=1}^p\ti x_k\zk\}
\ra}\\
\ds{
=(2\pi)^{-N/2}\la\exp\{i\sum_{\mu=1}^p(N^{-1/2}\sum_{k=1}^N\xm_k\zk)
(N^{-1/2}\sum_{j=1}^N\xm_j)\}\ra}\\
\ds{
=(2\pi)^{-N/2}\prod_{\mu=1}^p\la e^{i\ti u^\mu
\ti v^\mu}\ra},
\end{array}\label{G1}\end{equation}
where we use notations
\begin{equation}
\ti u^\mu\equiv N^{-1/2}\sum_{k=1}^N\xm_k\zk,\quad
\ti v^\mu\equiv N^{-1/2}\sum_{j=1}^N\xm_j.
\label{tiu}\end{equation}
It is easy to see that
\begin{equation}
\la e^{i\ti u^\mu\ti v^\mu}\ra=(2\pi)^{-1}\int du^\mu dv^\mu
\la e^{i(u^\mu\ti u^\mu+v^\mu\ti v^\mu)}\ra e^{-iu^\mu v^\mu}.
\label{G2}\end{equation}
Thus, using the inverse Fourier transform for the function
$F(\zeta_1,...,\zeta_N)$, we get
$$
\begin{array}{c}\ds{
P^g_N={1\over(2\pi)^{N/2}}\int\prod_{k=1}^N\theta(x_k-a_k)dx_k\int(\prod_{j=1}^N
d\zj)\exp\{-i\sum_{k=1}^Nx_k\zk\}F(\zeta_1,...,\zeta_N)}\\
\ds{
={1\over(2\pi)^{(N+p)}}\int(\prod_{\mu=1}^pe^{-iu^\mu v^\mu}du^\mu dv^\mu)\prod_{k=1}^N
\int dx_k\theta(x_k-a_k)\int d\zk
\la \exp\{-i\zk x_k}\\
\ds{ +\sum_{\mu=1}^p i(u^\mu\ti u^\mu+v^\mu\ti v^\mu)\}\ra =
{1\over (2\pi)^{N+p}}\int(\prod_{\mu=1}^pe^{-iu^\mu v^\mu}du^\mu dv^\mu)
\prod_{k=1}^N\int dx_k\theta(x_k-a_k) }\\
\ds{\cdot \int d\zk e^{-i\zk x_k}
\int(\prod_{\mu=1}^p{e^{-(\xm_k)^2/2}\over\sqrt{2\pi}})
\exp\{i(N^{-1/2}\sum_{\mu=1}^p u^\mu\xm_k\zk+
N^{-1/2}\sum_{\mu=1}^p v^\mu\xm_k)\} }
\end{array}$$
\begin{equation}\begin{array}{c}\ds{
={1\over(2\pi)^{(N+p)}}\int(\prod_{\mu=1}^pe^{-iu^\mu v^\mu}du^\mu dv^\mu)\prod_{k=1}^N
\int dx_k\theta(x_k-a_k)}\\
\ds{
\cdot\int d\zk
\cdot e^{-i\zk x_k}(\prod_{\mu=1}^p\exp\{-{(\um\zk+\vm)^2\over 2N}\})}\\
\ds{
={1\over (2\pi)^{({N\over 2}+p)}}\int(\prod_{\mu=1}^p\exp\{-iu^\mu v^\mu-
{(v^\mu)^2\over 2}\}du^\mu dv^\mu)}\\
\ds{
\cdot\prod_{k=1}^N
\int dx_k{\theta(x_k-a_k)\over U}
\exp\{{(ix_k+N^{-1}\sum_{\mu=1}^p u^\mu v^\mu)^2\over 2U^2}\} ,}
\end{array}
\label{G3}\end{equation}
where $U\equiv (N^{-1}\sum_{\mu=1}^p(u^\mu)^2)^{1/2}$.
Therefore we have
\begin{equation}\begin{array}{c}\ds{
P^g_N=(2\pi)^{-p}\int(\prod_{\mu=1}^pdu^\mu dv^\mu)
\exp\{-i\sum_{\mu=1}^pu^\mu v^\mu-{1\over
2}\sum_{\mu=1}^p(v^\mu)^2\}\cdot}\\
\ds{
\cdot\prod_{k=1}^N
H({a_k-iN^{-1}\sum_{\mu=1}^p u^\mu v^\mu\over U}).}
\end{array} \label{G5}\end{equation}
Now let us fix $\ov u=\{u^\mu\}_{\mu=1}^p$ and   change variables in the integral with
respect to $\ov v=\{v^\mu\}_{\mu=1}^p$,
\begin{equation}
v_1={1\over\sqrt N}(\ov e_1,\ov v),\,\, v_2=(\ov e_2,\ov v),...,v_p=(\ov e_p,\ov v),
\label{G6}\end{equation}
where $\{\ov e_i\}_{i=1}^p$ is the orthonormal system of vectors in
${\bf R}^p$ such that $e_1^\mu=( U\sqrt N)^{-1}u^\mu$. Then, integrating
with respect $v_2,...,v_p$, we obtain
\begin{equation}\begin{array}{c}\ds{
P^g_N=(2\pi)^{-(p-1)/2}\int(\prod_{\mu=1}^pdu^\mu)\int dv_1\exp\{-iNUv_1-
{N\over 2}(v_1)^2}\\
\ds{
+[N\delta]\log H({a_1\over U}-iv_1)+(N-[N\delta])
\log H({a_2\over U}-iv_1)\}.}
\end{array}\label{G7}\end{equation}
Using the spherical coordinates in the integral with
respect to $\ov u$ and integrating with respect to angular
variables, we get
\begin{equation}\begin{array}{c}\ds{
P^g_N=\Gamma(p)\int_0^\infty dU\int dv_1\exp\{(p-1)\log U-iNUv_1-
{N\over 2}(v_1)^2}\\
\ds{
+[N\delta]\log H({a_1\over U}-iv_1)+(N-[N\delta])\log H({a_2\over U}-iv_1)\}.}
\end{array}\label{G8}\end{equation}
Let $V(U)$ be the point of minimum with respect to $V$ of the
function ${\cal F}_0(U,V)$ defined by (\ref{F_0}). Let
us change the path of integration with respect to $v_1$ in
(\ref{G8}) from the real axis to the line $L$ which is parallel to
it, but contains the point $z=-iV(U)$. Then, following the
saddle point method, we divide the integral into two parts
\begin{equation}\begin{array}{c}\ds{
P^g_N=\Gamma(p)\int_0^\infty dU(\int_{|t|> N^{-1/3}}+
\int_{|t|\le N^{-1/3}}) dt\exp\{(p-1)\log U }\\
\ds{
-NUV(U)+{N\over 2}(V(U))^2-iNUt-
{N\over 2}t^2}\\
\ds{
+[N\delta]\log H({a_1\over U}-V(U)-it)+
(N-[N\delta])\log H({a_2\over U}-
V(U)-it)\}.}
\end{array}\label{G9}\end{equation}
Due to the simple inequality
\begin{equation}
|H(a+ic)|\le H(a)e^{c^2/2}, \label{boundH}\end{equation} valid for
any real numbers $a$ and $c$, we conclude, that the second
integral is $o(1)\exp\{N{\cal F}_0(U,V;\a,\delta,q,q')\}$. Replacing
in the first integral ${\cal F}_0(U,V(U)-it)$ by its Taylor
expansion up to the second order term (the first order term is
zero due to the choice $V(U)$) and then performing the Gaussian
integration, we see that
\begin{equation}
P^g_N\le\Gamma(p)\int_0^\infty dU\exp\{N({\cal
F}_0(U,V(U);\delta,q,q')+o(1))\} . \label{G10}\end{equation}
Applying the standard Laplace method, we conclude that    for the
Gaussian random variables $\xm_k$  Eq.  (\ref{mest}) can be
replaced by the following stronger statement:
$$
\limsup_{N\to
\infty}{1\over N}\log P^g_{N}= \max_{U>0}{\cal
F}_0(U,V(U);\delta,q,q')- {\alpha\over
2}\log\alpha+{\alpha\over 2}. $$
\medskip

The difference of non-Gaussian case from the Gaussian one is that
we have, in the sixth line of (\ref{G3}),
$\prod_{\mu=1}^p\cos{\um\zk+\vm\over\sqrt N}$ instead of
$\prod_{\mu=1}^p\exp\{-{(\um\zk+\vm)^2\over 2N}\}$. To replace the
former term by the latter one we have to  estimate  the difference
between them for different $\uv$, $\vv$ and $\ov\zeta$. To this
end we introduce some smoothing factors in the integration
(\ref{G3}).

\begin{lemma}\label{lem:1}
\begin{equation}
\la\prod_{k=1}^N\theta(\ti x_k-a_k)\ra\le
P_N^1l_N^{p/2}(1-e^{-h^2/2\l})^{-N}e^{No(1)}+e^{-\const N(\e_N^*)^{-1/2}},
\label{l1}\end{equation}
where
$$\begin{array}{c}\ds{
P_N^1\equiv
{1\over(2\pi)^{N+p}}\int d\uv
d\vv\exp\{-il_N(\uv,\vv)-\e_N^*{(\uv,\uv)\over 2N}-\e_N^*{(\vv,\vv)\over
2N}\}}\\
\ds{
\cdot \prod_{k=1}^N \int d\zk\hhN(\zk)e^{-\l\zk^2/2- ia_k\zk}
\prod_{\mu=1}^{p}\cos{\um\zk+\vm\over\sqrt N}},
\end{array}
$$
 $\e_N^*=(\log\log N)^{-1}$, $l_N\equiv{1\over 2}+{1\over 2}
\sqrt{1-4(\e_N^*)^2}$, $\l$ is a fixed positive number and $$
\hhN(\zeta)= {2\over \zeta}\sin\zeta{N^{1/2+d}+2h\over
2}\exp\{-i\zeta{N^{1/2+d}\over 2}\} $$ is the complex conjugate of
 the Fourier transform of $\hN(x)$  -the  characteristic function of
the interval $(-h,N^{1/2+d}+h)$ with some positive $d$ and
$h>({2\lambda\over \pi})^2$. Here and below $\vv=(v^1,\dots,v^p)$,
$\uv=(u^1,\dots,u^p)$, $d\vv=\prod_{\mu=1}^p d\vm$ and
$d\uv=\prod_{\mu=1}^p d\um$.
\end{lemma}

\begin{remark}
\no In fact we can take $\e_N^*\to 0$ as slowly as we want,
we can even fix $\e_N^*=\e$ with $\e$ being
 small enough. However, in this case we have to be more careful
 to control the constants which will appear in our  estimates.
\end{remark}
\medskip

Now we start to prove Theorem \ref{thm:1}. Denote
\begin{equation}\begin{array}{c}\ds{
F_{N,k}(\uv,\vv)={1\over 2\pi}\int d\zk\hhN(\zk)e^{-\l\zk^2/2-
ia_k\zk} \prod_{\mu=1}^{p}\cos{\um\zk+\vm\over\sqrt N};}\\
\ds{
 \ti F(\uv,\vv)=\prod_{k}F_{N,k}(\uv,\vv). }
\end{array}\label{1}\end{equation}
To simplify formulae  in the places where it is not important,
 we confine ourselves to the case $a_k=a$. Since in this case all
$F_{N,k}(\uv,\vv)$ are identical, we could omit
the index $k$.

To replace  the product term of $cos$ in Eq. (\ref{1}) by the
exponent we modify a method originally proposed by Lyapunov. He
employed it to  prove that the distribution of the sum of
independent variables uniformly converges  to the normal
distribution (see \cite{Loe}). To ensure the  method  to work, the
second and the third moments of the random variables must be
bounded. Since in our setting the random variables have the form
$\um\xm_k$ and $\vm\xm_k$ and  their  moments coincide with
$|\um|^{2,3}$ and $|\vm|^{2,3}$,  we need to remove  large $|\um|$
and $|\vn|$ in the integrals. For this purpose  we
 take $\e_N=(\log N)^{-1}$  and
denote
\begin{equation}
\chi_{\e_N}(\um,\vm)=\theta (\e_N^2\sqrt N-|\um|)\theta (\e_N\sqrt
N-|\vm|) . \label{1.1}\end{equation}

Note that the different powers of $\e_N$ in the $\theta$-functions for
$u$ and $v$ are necessary in our estimates below.

Rewrite
\begin{equation}\begin{array}{c}\ds{
P_N^1={1\over (2\pi)^p}\int e^{-il_N(\uv,\vv)}\ti
F(\uv,\vv)\exp\{-{\e_N^*\over 2}(\vv,\vv)- {\e_N^*\over
2}(\uv,\uv)\} d\uv d\vv}\\
\ds{
=\sum_{m=0}^pC_p^m\int d\uv
d\vv\prod_{\mu=1}^m(1-\chi_{\e_N}(\um,\vm))
\prod_{\nu=m+1}^p\chi_{\e_N}(\un,\vn)}\\
\ds{
\cdot e^{-{ \e_N^*\over 2}(\uv,\uv)} e^{-{\e_N^*\over2}(\vv,\vv)}
 e^{-il_N(\uv,\vv)}\ti F(\uv,\vv)\equiv \sum_{m=0}^pC_p^m I_m.}
\end{array} \label{2}\end{equation}
Let us first estimate $I_m$ in the above equation
\begin{equation}\begin{array}{c}\ds{
|I_{m}|\le {1\over (2\pi)^p}\int d\uv
d\vv\prod_{\mu=1}^m(1-\chi_{\e_N}(\um,\vm)) }\\
\ds{
\cdot\prod_{\nu=m+1}^p\chi_{\e_N}(\un,\vn)e^{-{\e_N^*\over 2}(\uv,\uv)}
 e^{-{\e_N^*\over 2}(\vv,\vv)}
\prod_{k=1}^N\int d\zk|\hhN(\zk)|e^{-\l\zk^2/2}}.
\end{array}\label{3}\end{equation}
Now, using the bound
\begin{equation} \begin{array}{c}\ds{
\int d\zk|\hhN(\zk)|e^{-\l\zk^2/2}}\\
\ds{
 =\int
d\zk|{2\over\zk}\sin(\zk{N^{1/2+d} \over 2})|
e^{-\l\zk^2/2}\le\const\log N,}
\end{array}\label{4}\end{equation}
we arrive at
$$
|I_m|\le e^{\const N\log\log
N}(\e_N^*)^{-p}e^{-mN\e_N^*\e_N^4/2}. $$ Thus,
\begin{equation}
|\sum_{m=m_0}^pC_p^mI_m|\le e^{-\const N\log\log N},
\label{5}\end{equation}
where $m_0=[(\log N)^{5}] >>(\e_N^*)^{-1}\e_N^{-4}\log\log
N$.

In the following  it would be more convenient to
have the integration with respect to $u^1,\dots,u^m$ and
$v^1,\dots,v^m$ in the whole ${\bf R}$. Therefore, we
perform the first product in (\ref{2}) and rewrite
$\sum_{m=0}^{m_0}C_p^mI_m$ in the form

\begin{equation}
\sum_{m=0}^{m_0}C_p^mI_m=\sum_{m=0}^{m_0}\ti C_m\ti I_m,
\label{5.1}\end{equation}
where
\begin{equation}\begin{array}{c}\ds{
\ti I_m\equiv {1\over (2\pi)^p}\int d\uv
d\vv\prod_{\nu=m+1}^p\chi_{\e_N}(\um,\vm)
e^{-{\e_N^*\over 2}(\uv,\uv)} e^{-{\e_N^*\over 2}(\vv,\vv)}
 e^{-il_N(\uv,\vv)}\ti F(\uv,\vv)}
\end{array}\label{5.2}\end{equation}
and $\ti C_m$ are some combinatorial coefficients.
 These coefficients are not important, because for
our choice of $m$ ($m\le m_0=o(N)$) all of them are of the order
$e^{o(N)}$ and after taking the logarithm and dividing by $N$
give us $o(1)$-terms.
Thus, we have
\begin{equation}
P_N^1=\sum_{m=0}^{m_0}\ti C_m\ti I_m+ O(e^{-\const N\log\log N}).
\label{6}\end{equation}
\medskip

 To proceed further we define
\begin{equation}\begin{array}{c}\ds{
F^{(m)}(\uv_1,\vv_1;\uv_2,\vv_2)\equiv \exp\{-{(\vv_2,\vv_2)\over 2N}\}}\\
\ds{
\cdot\la
H_{N,h,\ti U}({a-h-i{(\uv_2,\vv_2)\over N}-{(\uv_1,\ov\xi_1)\over
\sqrt N}
\over\sqrt{\ti U^2+\l}})\exp\{i{(\vv_1,\ov\xi_1)\over\sqrt N}\}\ra
}\\
\ds{
=\int d\zeta\hhN(\zeta)e^{-\l\zeta^2/2- ia\zeta}\prod_{\mu=1}^m\cos
{\um\zeta+\vm\over\sqrt N}}\\
\ds{  \exp\{-{1\over 2N}(\vv_2,\vv_2)-
{1\over N}(\uv_2,\vv_2)\zeta-{1\over 2N}(\uv_2,\uv_2)\zeta^2\},}
\end{array}\label{9}\end{equation}
where
\begin{equation}
H_{N,h,\ti U}(x)={1\over\sqrt{2\pi}}\int_0^\infty
\theta({N^{1/2+d}+2h\over\sqrt{\ti U^2+\l}}-t)\exp\{-{1\over 2}
(t+x)^2\}dt.
\label{9'}\end{equation}
Here
and  below $\uv_1\equiv(u^{1},...,u^m)$ and $\vv_1\equiv
(v^{1},...,v^m)$, $\uv_2\equiv(u^{m+1},...,u^p)$, $\vv_2\equiv
(v^{m+1},...,v^p)$, so that $\uv=\{\uv_1,\uv_2\}$,
$\vv=\{\uv_1,\uv_2\}$,  $\ov\xi_1\equiv (\xi^{1}_1,...,\xi^m_1)$ is
the random vector with independent components, assuming values $\pm 1$
with probability ${1\over 2}$,
$\la...\ra$ means the average with respect to $\ov\xi_1$
and $\ti U\equiv[{1\over N}(\uv_2,\uv_2)]^{1/2}$.
Expression (\ref{9}) is obtained from (\ref{1}) by changing
$cos$ in the product $\prod_{\mu=m+1}^p$ by the correspondent exponent and
then by integration with respect to $\zk$.

The main technical tool at this step is a lemma, which is a modification
of the Lyapunov theorem.

\begin{lemma} \label{lem:2}
For any $\uv_2,\vv_2,\lv_2$ such that
$|\un|,|\vn|,|\ln|\le\e_N\sqrt
N$ and any $\uv_1,\vv_1,\lv_1$ the function
 $$
R^{(m)}(\uv_1,\wv_1;\uv_2,\wv_2)\equiv
F_N(\uv_1,\wv_1;\uv_2,\wv_2)-F^{(m)}(\uv_1,\wv_1;\uv_2,\wv_2)
$$
admits the bound
\begin{equation}\begin{array}{c}\ds{
|R^{(m)}(\uv_1,\wv_1;\uv_2,\wv_2)|\le
\const\e_N^2(1+{(\lv_2,\lv_2)\over N})(\ti U^2+
\l)^{1/2}}\\
\ds{
\cdot\exp\{-{\l(\vv_2,\vv_2)\over
4N(\ti U^2+\l)}+{(\lv,\lv)\over N}\}+
 \exp\{-\const\e_N^{-4}+{(\lv,\lv)\over N}\}. }
\end{array}\label{11}\end{equation}
Here and below $\wv\equiv\vv+i\lv$.
\end{lemma}

\medskip

This lemma allows us to replace in our formulae $F_{N}$ by $F^{(m)}$
in the following sense. Let us write
\begin{equation}\begin{array}{c}\ds{
\ti I_{m}\equiv {1\over (2\pi)^p}\int d\uv d\vv
\prod_{\nu=m+1}^p\chi_{\e_N}(\um,\vm)
\exp\{-il_N(\uv,\vv)\}}\\
\ds{
\cdot(F^{(m)}(\uv_1,\vv_1,\uv,\vv)+
R^{(m)}(\uv_1,\vv_1,\uv_2,\vv_2))^{N}
e^{-{\e_N^*\over 2}(\vv,\vv)}e^{-{\e_N^*\over 2}(\uv,\uv)}}\\
\ds{
\equiv\sum^{N}_{k=0}C_{N}^{k}I_{m,k}}\,\, ,
\end{array}\label{12}\end{equation}
where
$$\begin{array}{c}\ds{
I_{m,k}\equiv {1\over (2\pi)^p}\int d\uv d\vv
\prod_{\nu=m+1}^p\chi_{\e_N}(\um,\vm)
e^{-il_N(\uv,\vv)}(F^{(m)}(\uv_1,\vv_1,\uv,\vv))^{N-k}}\\
\ds{
\cdot(R^{(m)}(\uv_1,\vv_1,\uv_2,\vv_2))^{k} e^{-{\e_N^*\over 2}(\vv,\vv)}
e^{-{\e_N^*\over 2}(\uv,\uv)}.}
\end{array}$$
\begin{lemma}\label{lem:3}
For $k>k_0\equiv[N\log^{-1/2}\e_N^{-1}]$
$$\begin{array}{c}\ds{
|I_{m,k}|\le e^{N\const}(\e_N)^{2k}(\e_N^*)^{-2p}
\exp\{-k\const \log\e_N^{-1}\}.}
\end{array} $$
 \end{lemma}
Thus, we get that for $k>k_0$ $I_{m,k}$ have the
order $e^{-N\const\log^{1/2}\e_N^{-1}}$ and so we can neglect these terms
in (\ref{12}).

\medskip

Now we shall study the leading terms in the r.h.s. of Eq. (\ref{12})
($I_{m,k}$ with $k<k_0$). In fact,
the next step is  a version of the saddle point method
(cf.(\ref{G8})-(\ref{G10})).

Let us take any real fixed $V$ and change the path of integration
w.r. to $\vv_2$
  from the product of intervals
$(-\e_N\sqrt N,\e_N\sqrt N)$ to the product of the paths
$L^\nu_1\cup L^\nu_2$, with $L^\nu_1=(-\e_N\sqrt N-{iV\un\over
\ti U},\e_N\sqrt N-{iV\un\over \ti U})$ and $L^\nu_2= (-\e_N\sqrt
N,-\e_N\sqrt N-{iV\un\over \ti U })\cup (\e_N\sqrt
N-{iV\un\over \ti U},\e_N\sqrt N)$ ($\nu=m+1,...N$). It can be
done, since all our functions are analytical w.r.to $\vn$,

Then take any real  $\lm$, such that $(\lv_1,\lv_1)\le N\const$
and choose the paths of integration with respect to $\vv_1$ as
$L^\mu=\{\wm=\tm-i\lm,\tm\in{\bf R}\}$. Finally, we get
\begin{equation}\begin{array}{c}\ds{
I_{m,k}={1\over (2\pi)^p}\sum_{n=1}^{p-m}C_{p-m}^n\int d\uv_1
\int_{\prod_{\mu=1}^m L^\mu}d\wv_1 \int_{\prod_{\nu=m+1}^{p-n}
L^\nu_1}d\wv_3}\\
\ds{
\cdot\int_{\prod_{\nu=p-n+1}^{p}L^\nu_2}d\wv_4
\int_{-\e_N^2\sqrt N}^{\e_N^2\sqrt N}d\uv_3 d\uv_4
e^{-\e_N^*(\wv,\wv)/2}e^{-\e_N^*(\uv,\uv)/2}}\\
\ds{
\cdot e^{-il_N(\uv,\wv)}
(F^{(m)}(\uv,\wv))^{N-k}
(R^{(m)}(\uv,\wv))^{k}\equiv
 \sum_{n=1}^{p-m}C_{p-m}^n I_{m,k,n}}.
 \end{array}\label{14}\end{equation}
Here and below $\uv=\{\uv_1,\uv_3,\uv_4\}$,
$\wv=\{\wv_1,\wv_3,\wv_4\}$, where $\uv_1,\wv_1$ are the same as
before and we divide vectors $\wv_2$ and $\uv_2$ in two sub-vectors
$\uv_2=\{\uv_3,\uv_4\}$, $\wv_2=\{\wv_3,\wv_4\}$ in such a way
that $\uv_4,\wv_4$ include the last $n$ components of
$\uv_2$ and $\wv_2$ respectively.

Now let us get rid of $I_{m,k,n}$ with sufficiently large $n$. Similarly
to the proof of Lemma \ref{lem:3} on the basis of Lemma \ref{lem:2}, we get
\begin{equation}
|I_{m,k,n}|\le
e^{N\const}(\e_N^*)^{-p}e^{-\const nN\e_N^2}
\exp\{(\lv_1,\lv_1)+NV^2\}.
\label{15}\end{equation}
So, taking $n>n_0=[\e_N^{-5/2}]$, on the basis of (\ref{15}) one can
conclude that we need to study only the first $n_0$ terms in
(\ref{14}).

\medskip

We remark that starting from this moment, we shall distinguish the
terms with $a_1$ and $a_2$. Denote
\begin{equation}\begin{array}{c}\ds{
G_m^*(U,V,\uv_1,\lv_1)\equiv }\\
\ds{
\la H({a_1-h-VU-
N^{-1/2}(\uv_1,\ov\xi_1)\over\sqrt{U^2+\l}})
\exp\{{(\lv_1,\ov\xi_1)\over\sqrt N}\}\ra^\delta }\\
\ds{
\cdot\la H({a_2-h-VU-
N^{-1/2}(\uv_1,\ov\xi_1)\over\sqrt{U^2+\l}})
\exp\{{(\lv_1,\ov\xi_1)\over\sqrt N}\}\ra^{1-\delta}}\\
\ds{
\cdot\exp\{-{l_N\over N}(\uv_1,\lv_1)-l_NUV+{1\over 2}V^2\} .}
\end{array}\label{16}\end{equation}

\begin{lemma}\label{lem:4}
Let  $G_{m,k,n}(V,\uv_1,\lv_1,\uv_3)$ be the function
which we get, if in (\ref{14}) integrate with respect to
 $\wv_1$, $\wv_3$, $\uv_4$ and $\wv_4$. Then
\begin{equation}\begin{array}{c}\ds{
|G_{m,k,n}(V,\uv_1,\lv_1,\uv_3)|}\\
\ds{
\le(2\pi)^{-p/2}(G_m^*(U,V,\uv_1,\lv_1))^Ne^{-{\e_N^*\over 2}(\uv_1,\uv_1)
+No(1)}.}
\end{array}\label{17}\end{equation}
Here and below $U=[N^{-1}(\uv_3,\uv_3)]^{1/2}$, so that
$\ti U^2=U^2+N^{-1}(\uv_4,\uv_4)$.
\end{lemma}

\medskip
 Once we have an upper bound for $G_{m,k,n}$ we can estimate all the $\ti I_m$
in (\ref{6}).
Let us study first the term  with $m=0$.
Consider the function
\begin{equation}\begin{array}{c}\ds{
{\cal F}_{\l,h}(U,V)\equiv\delta\log
H({a_1^*-h-V U\over\sqrt{U^2+\l}}
 )+(1-\delta)\log H({a_2^*-h-V U\over\sqrt{U^2+\l}})}\\
\ds{
-UV+{1\over 2}V^2}.
\end{array}\label{22}\end{equation}
Let $ V(U)$ be chosen  from the condition
\begin{equation}
{\cal F}_0(U,V(U);\a,\delta,q,q')=\min_V{\cal F}_0(U,V;\a\delta,q,q').
\label{V(U)}\end{equation}

The function  ${\cal F}_{\l,h}(U,V(U))$ and the functions which appear in the
exponent of (\ref{17}) for $m=0$ satisfy the inequalities of the type
$$
{\cal F}_{\l,h}(U,V(U))\le \alpha\log U-
{U^2\over 2}
$$
(it follows from  $\log H(x)\le 0$ and $V(U)\le U$).
Thus, since
$a_{1,2}\to a_{1,2}^*$ and $l_N\to 1$ as $N\to \infty$, on the basis of
 (\ref{17}) for $m=0$, we get
$$
|\ti I_0|\le (2\pi)^{-p/2}\int d\uv_3\exp\{N[{\cal F}_{\l,h}(U,
V(U))+o(1)]\}, $$
where $\ti I_0$ is defined by formula (\ref{5.2}) for $m=0$.
\begin{remark}\label{rem:L}
Let us note that here we have use the following simple statement.
 If the continuous functions $\phi(U)$, $\phi_N(U)$ ($N=1,2,..$)
 ($U\in{\bf R}_+$) satisfy the inequalities
\begin{equation}\begin{array}{c}
\phi(U),\phi_N(U)\le -C_1U^2,\quad U\ge L,\\
\phi(U),\phi_N(U)\le C_2\log U,\quad U\le \e,
\end{array}\label{rem.L}\end{equation}
with some positive $C_1$ and $C_2$ and $\phi_N(U)\to \phi(U)$, as $N\to\infty$, uniformly in
each compact set in ${\bf R}_+$, then $\int\exp\{N\phi_N(U)\}dU=$
$e^{o(N)}\int\exp\{N\phi(U)\}dU$.
The proof of this statement is very simple, and we omit it.
\end{remark}
Below we shall use this remark without additional comments.

Performing the  spherical change of variables and using  the
Laplace method, we get now

\begin{equation}
|\ti I_0|\le\exp\{
N[\max_{U}{\cal F}_{\l,h}(U, V(U))+\alpha\log U-{\alpha\over 2}
\log\alpha+{\alpha\over 2}+o(1)]\} .
\label{23}\end{equation}

\bigskip

To study the terms with $m\not=0$ we  chose $\lv_1(U,V,\uv_1)$ in
such a way that
\begin{equation}
G_m^*(U,V,\uv_1,\lv_1(U,V,\uv_1))=\min_{\lv_1\in{\bf
R^m}}G_m^*(U,V,\uv_1,\lv_1),
\label{20}\end{equation}
where the function $G_m^*$ is defined by (\ref{16}).
Then we use
 the inequality, which follows from the
fact that $(\log H(x))''\le 0$,
\begin{equation}
H(x+y)\le H(x) e^{-A(x)y}
\label{27}\end{equation}
with the function $A(x)$ defined by (\ref{A(x)}).
On the basis of this inequality we get
\begin{equation}\begin{array}{c}\ds{
\la H({a_{1,2}-h-VU-
N^{-1/2}\sum_{\mu=1}^m\um\xi^\mu_1\over\sqrt{U^2+\l}})
\exp\{\sum_{\mu=1}^m\lm{\xi^\mu_1\over\sqrt
N}\}\ra}\\
\ds{
 \le\la H({a_{1,2}-h-VU\over\sqrt{U^2+\l}})
\exp\{\sum_{\mu=1}^m(A_{1,2}^{(\l,h)}\um+\lm){\xi^\mu_1\over\sqrt
N}\}\ra}\\
\ds{
= H({a_{1,2}-h-VU\over\sqrt{U^2+\l}})\prod_{\mu=1}^m
\cosh{A_{1,2}^{(\l,h)}\um+\lm\over\sqrt N}} \\
\ds{
\le H({a_{1,2}-h-VU\over\sqrt{U^2+\l}})
\exp\{{1\over 2N}\sum_{\mu=1}^m(A_{1,2}^{(\l,h)}\um+\lm)^2\}, }
\end{array}\label{27a}\end{equation}
where
\begin{equation}
A_{1,2}^{(\l,h)}=(U^2+\l)^{-1/2}A({a_{1,2}-h-VU\over\sqrt{U^2+\l}}) .
\label{27b}\end{equation}
Thus,
\begin{equation}\begin{array}{c}\ds{
G_{m}^*(U,V,\uv_1,\lv_1(U,V,\uv_1))|\le
\exp\{\delta\log
H({a_1-h-UV\over  \sqrt{U^2+\l}}) }\\
\ds{
+(1-\delta)\log H({a_2-h-UV\over
\sqrt{U^2+\l}})-l_NUV+{1\over 2}V^2}\\
\ds{
+\min_{\lm}[{\delta\over 2N}\sum_{\mu=1}^m(A_{1}^{(\l,h)}\um+\lm)^2
+{1-\delta\over 2N}\sum_{\mu=1}^m(A_{2}^{(\l,h)}\um+\lm)^2}\\
\ds{
-{l_N\over N}
\sum_{\mu=1}^m\lm\um]\}.}
\end{array}\label{28}\end{equation}
Taking $\lm=(1-A_1^{(\l,h)}\delta-A_2^{(\l,h)}(1-\delta))\um$, which give us
the minimum of the
expression in the r.h.s. of (\ref{28}), we get
\begin{equation}\begin{array}{c}\ds{
|G_{m}^*(U,V,\uv_1,\lv_1(U,V,\uv_1))|\le
\exp\{N[\delta\log
H({a_1-h-UV\over  \sqrt{U^2+\l}})}\\
\ds{
+(1-\delta)\log H({a_2-h-UV\over
\sqrt{U^2+\l}})
-UV+{1\over 2}V^2]-D^{(\l,h)}(U,V)(\uv_1,\uv_1)\},}
\end{array}\label{28a}\end{equation}
where $D^{(\l,h)}(U,V)$ is defined by (\ref{D(U)}) if we substitute
there $A_{1,2}(U,V)$ by $A_{1,2}^{(\l,h)}(U,V)$. From (\ref{28a}) it is easy to see that if $D^{(\l,h)}(U,V)\ge 0$, then
\begin{equation}\begin{array}{c}\ds{
|\int d\uv_1G_{m}^*(U,V,\uv_1,\lv_1(U,V,\uv_1))\exp\{-{\e_N^*\over 2}
(\uv_1,\uv_1)\}|}\\
\ds{
\le e^{No(1)}\exp\{N[\delta\log
H({a_1^*-h-UV\over  \sqrt{U^2+\l}})+
(1-\delta)\log H({a_2^*-h
-UV\over
\sqrt{U^2+\l}})}\\
\ds{
-UV+{1\over 2}V^2]\}.}
\end{array}\label{28.1}\end{equation}
If $D^{(\l,h)}(U,V)$ is negative, we use
\begin{proposition} \label{pro:1}
If $D^{(\l,h)}(U,V)<0$, $\l$ and $h$ are small enough, then
\begin{equation}\begin{array}{c}\ds{
|\int d\uv_1G_{m}^*(U,V,\uv_1,\lv_1(U,V,\uv_1))\exp\{-{\e_N^*\over 2}
(\uv_1,\uv_1)\}|}\\
\ds{
\le\exp\{N[{\delta\over 1-2D^{(\l,h)}(U,V)}\log
H({a_1^*-h-UV\over  \sqrt{U^2+\l}}) }\\
\ds{
+{1-\delta\over 1-2D^{(\l,h)}(U,V)}\log H({a_2^*-h-UV\over
\sqrt{U^2+\l}})-UV+{1\over 2}V^2+o(1)]\}.}
\end{array}\label{28b}\end{equation}
\end{proposition}

Thus, on the basis of (\ref{28a}) and (\ref{28b}), we have got
that for any $n$-independent finite $V$,
$$\begin{array}{c}\ds{
 |\int d\uv_1
G_{m}^*(U,V,\uv_1,\lv_1(U,V,\uv_1))\exp\{-{\e_N^*\over 2}
((\uv_1,\uv_1)+(\uv_3,\uv_3))\}|}\\
\ds{
 \le\exp\{N[{\cal F}_{\l,h}^{D}(U,V)+o(1)]\},}
\end{array}$$
where  for $D^{(\l,h)}(U,V)<
0,$
 ${\cal F}_{\l,h}^{D}(U,V)$ is defined by the expression
in the exponent in the r.h.s. of (\ref{28b}) and for
$D^{(\l,h)}(U,V)\ge 0,$ it coincides with ${\cal
F}_{\l,h}(U,V)$. Then, choosing $V$ to minimise this
estimate for any $U$, we get
\begin{equation}\begin{array}{c}\ds{
\int d\uv_3|\int d\uv_1
G_{m}^*(U,V,\uv_1,\lv_1(U,V,\uv_1))\exp\{-{\e_N^*\over 2}
((\uv_1,\uv_1)+(\uv_3,\uv_3))\}|}\\
\ds{
\le\int dU
\exp\{N[\min_V{\cal F}_{\l,h}^{D}(U,V)+\alpha\log U
-{\alpha\over 2}\log\alpha+ {\alpha\over 2}+o(1)]\}.}
\end{array}\label{33}\end{equation}
  Thus,  for any $m\le m_0=o(N)$,
$$
|\ti I_m|\le \exp\{N[
\max_U\{\min_V{\cal F}_{\l,h}^D(U,V)+\alpha\log
U\}-{\alpha\over 2}\log\alpha+ {\alpha\over 2}+o(1)]\}.
$$
 Hence,
$$
P_N\le \exp\{N[
\max_U\min_V\{{\cal F}_{\l,h}^D(U,V)+\alpha\log
U\}-{\alpha\over 2}\log\alpha+ {\alpha\over 2}+o(1)]\}.
$$
Therefore, on the basis of Lemma \ref{lem:1}, we have
$$\begin{array}{c}\ds{
\limsup_{N\to \infty}{1\over
N}\log\la\prod_{k=1}^{[\delta N]}\theta(\ti x_k-a_1)
\prod_{k=1+[\delta N]}^N\theta(\ti x_k-a_2)\ra }\\
\ds{
\le\max_U\min_V\{{\cal F}_{\l,h}^D(U,V)\}+\alpha\log
U\}-{\alpha\over 2}\log\alpha+ {\alpha\over 2}+o(1).}
\end{array}$$
We get the conclusions of Theorem~\ref{thm:1}, after taking the limits
$\lambda\to 0$ and then $h\to 0$.

\bigskip

\no {\it Proof of Theorem \ref{thm:2}.}
 To prove Theorem \ref{thm:2} let us show that if $\alpha$ is small
enough to satisfy the condition
\begin{equation}
 e^{-{1\over 2\alpha}}<\alpha^4,
\label{conda'}\end{equation}
then
\begin{equation}\begin{array}{c}\ds{
\max_U\min_V {\cal F}_0^D(U,V;\a,\delta,0,0)\le \log
H({a_1^*-\alpha\over\sqrt\alpha})+(1-\delta) \log
H({a_2^*-\alpha\over\sqrt\alpha})}\\
\ds{
+ {\alpha\over
2}\log\alpha- {\alpha\over
2}+O(\delta^2\alpha^{-3})+O(e^{-1/\alpha})}\\
\ds{
= {\cal F}_0(\sqrt\alpha,\sqrt\alpha;\a,\delta,0,0)
+O(\delta^2\alpha^{-3})+O(e^{-1/\alpha}).}
\end{array}\label{F(a)}\end{equation}
By  virtue of the condition $\delta<<\alpha^3\log\alpha^{-1}$, we
get then the statement (\ref{esta}) of Theorem \ref{thm:2}.

We start, proving (\ref{F(a)}) for $U>2\sqrt\alpha$.
\begin{proposition}\label{pro:2}

If $U>2\sqrt\alpha$, and $V(U)$ is defined by condition
(\ref{V(U)}), then $\sqrt\alpha\le V(U)\le U$.
\end{proposition}

 On the basis of Proposition \ref{pro:2}, we get
\begin{equation}\begin{array}{c}\ds{
{\cal F}_0^D(U,V(U);\a,\delta, 0, 0)\le
\alpha\log U-V(U)U+{1\over 2}(V(U))^2}\\
\ds{
\le\alpha\log U-\sqrt\alpha U+{\alpha\over 2}\le
{\alpha}\log 2\sqrt\alpha- 2\alpha+{\alpha\over 2}.}
\end{array}\label{T2.1}\end{equation}
Here the first inequality is due to $\log H(x)\le 0$,
while the second and the third follow from Proposition \ref{pro:2}.
But, using the asymptotic formulae
\begin{equation}\begin{array}{c}\ds{
H(x)={1\over x\sqrt{2\pi}}e^{-x^2/2}(1+O(1/x^2))\,\, (x>>1), }\\
\ds{
H(x)=1+{1\over x\sqrt{2\pi}}e^{-x^2/2}(1+O(1/x^2))\,\, (x<<-1),}
\end{array}\label{asH}\end{equation}
 and condition $\delta<<\alpha^3\log\alpha^{-1}$, it is
easy to get that the r.h.s. of (\ref{F(a)}) is
\begin{equation}
{\cal F}_0(\sqrt\alpha,\sqrt\alpha;\a,\delta,0,0)
\sim \alpha\log\sqrt\alpha-{\alpha\over 2}+o(\alpha^2)>
{\alpha}\log 2\sqrt\alpha- 2\alpha+{\alpha\over 2}.
\label{T2.2}\end{equation}
This inequality and (\ref{T2.1}) prove (\ref{F(a)}) for
$U>2\sqrt\alpha$.

Now let us check (\ref{F(a)}) for $U<0.5\sqrt\alpha$.
To this end let us write an equation for $V(U)$ which follows
from (\ref{V(U)}),
\begin{equation}
U=V+\delta A({\alpha+1-2\delta\over U}-V)+
(1-\delta) A({\alpha-(1-2\delta)\over U}-V),
\label{eqV}\end{equation}
where the function $A(x)$ is defined by (\ref{A(x)}). By using asymptotic formulae
\begin{equation}
A(x)=x(1+O({1\over x}))\,\, (x>>1), \,\,
A(x)={e^{-x^2/2}\over\sqrt{2\pi}}(1+O({1\over x}))\,\, (x<<-1),
\label{asA}\end{equation}
we get that in this case
$$
V(U)=U+ o(\alpha^2).
$$
Therefore
\begin{equation}\begin{array}{c}\ds{
{\cal F}_0^D(U,V(U);\a,\delta,0,0)\le
\alpha\log U-V(U)U+{1\over 2}(V(U))^2}\\
\ds{
\le\alpha\log U- {U^2\over 2}\le
{\alpha}\log 0.5\sqrt\alpha- {\alpha\over 8}.}
\end{array}\label{T2.3}\end{equation}
Now, using again  (\ref{T2.2}), we obtain
(\ref{F(a)}) for $U\le 0.5\sqrt\alpha$.

Now we are left to prove (\ref{F(a)}) for $0.5\sqrt\alpha
\le U\le 2\sqrt\alpha$. Let us prove first that for those
$U$ the function $D(U,V(U))$ defined by (\ref{D(U)}) is positive.
 To this end we use again asymptotic formulae (\ref{asA}).
Then we get
$$ \begin{array}{c}\ds{
A_1(U,V(U))=U^{-2}+o(\alpha^2)=O(\alpha^{-1}),\quad }\\
\ds{
A_2(U,V(U))=O(\alpha^{-1/2}e^{-1/8\alpha})=O(\sqrt\alpha).}
\end{array}$$
Here in the last equality we have used (\ref{conda'}).
 Using these
estimates, it is easy to obtain that $D(U,V(U))>0$ and therefore
for  $0.5\sqrt\alpha\le U\le 2\sqrt\alpha,$
$$
{\cal F}_0^D(U,V(U);\a,\delta, 0,0)= \min_V{\cal F}_0(U,V;\a,\delta,0,0).
$$
But
\begin{equation}\begin{array}{c}\ds{
\max_U\min_V {\cal F}_0(U,V;\a,\delta,0,0)\le
\max_U{\cal F}_0(U,U;\a,\delta,0,0)}\\
\ds{
=\max_U\{\alpha\log U-{U^2\over 2}
+\delta\log H({a_1^*\over U}-U)+(1-\delta)\log H({a_2^*\over U}-U)\}.}
\end{array}\label{F(U,U)}\end{equation}
Taking the derivative of the r.h.s. of (\ref{F(U,U)})
 with respect to $U$ we get:
\begin{equation} \begin{array}{c}\ds{
{\partial\over \partial U}{\cal F}_0(U,U;\a,\delta,0,0)= }\\
\ds{
{\alpha\over U}-U+\delta({a_1^*\over U^2}+1)A({a_1^*\over U}-U)+
(1-\delta)({a_2^*\over U^2}+1)A({a_2^*\over U}-U). }
\end{array}\label{derU}\end{equation}
Using  asymptotic formulae (\ref{asA}) we get
the equation for $U^*$ which is the maximum point of the r.h.s. of (\ref{F(U,U)}):
$$
{\alpha\over U^*}-U^*+O({\delta\over\alpha^{3/2}})+O(e^{-1/2\alpha})=0,
$$
so
$$
U^*=\sqrt\alpha+O({\delta\over\alpha^{3/2}})+O(e^{-1/2\alpha}).
$$
But since $\ds{ {d\over dU}(\alpha\log U- {1\over 2}U^2)
\bigg|_{U=\sqrt\alpha}=0}$, the Taylor expansion for this
function starts from the term $(U-\sqrt\alpha)^2$ and we get
$$
{\cal F}_0(U^*,U^*;\a,\delta,0,0)=
{\cal F}_0(\sqrt\alpha,\sqrt\alpha;\a,\delta,0,0)+
O(\delta^2\alpha^{-3})+O(e^{-1/\alpha}).
$$
Hence, we have proved (\ref{F(a)})
 and so (\ref{esta})  is proven.

Now one can  easily derive the estimate for $P_N^*(\delta,\alpha)$
from the inequality $$ P_N^*(\delta,\alpha)\le C_N^{[\delta N]}
P_N(0,0), $$ where $P_N(q,q')$ is defined by (\ref{P(q,q')}).
Thus, we have finished the proof of Theorem~\ref{thm:2}.
\medskip

\no{\it Proof of Theorem \ref{thm:3}.}
 It is easy to see that if for some $\e>0$ for any local minimum point
$\bs^*$ in $\Od^{1}$, we can find a point $\ov\s^{**}$ inside the ball $B_\delta^{1}$, 
such that
\begin{equation}
{\cal H}(\bs^*)-{\cal H}(\bs^{**})\ge\e^2 N,
\label{T3.1}\end{equation} 
then the event ${\cal A}$ takes place.
Let $\{x^*_k\}_{k=1}^N$   be the effective field generated
by the configuration $\bs^*$. Consider $I(\bs^*)\subset\{1,2,\dots,N\}$
- the set of indexes $i_1,\dots,i_{[N\delta]}$ such that
$\sigma^*_i\ti\xi^1_i=-1$. Assume that the
number $N_\e$ of indexes $i\in I(\bs^*)$ for which $x^*_k\le
-({1\over 2}+\alpha)\e$, is larger than $\e N$ (we denote the set
of these indexes by $ I_\e(\bs^*)$). Then consider the point
$\bs^{**}$, which differ from $\bs^*$ in the components with $[\e
N]+1$ first indexes $i\in I_\e(\bs^*)$, and coincides with $\bs^*$
in all the other components. Since we have changed only the
components of $\bs^*$ with indexes $i\in I_\e(\bs^*)\subset
I(\bs^*)$, $\bs^{**}\in B_\delta^{1}$. On the other hand,
\begin{equation}\begin{array}{c}\ds{
{\cal H}(\bs^*)-{\cal H}(\bs^{**})={1\over 2}(\ti{\bf J}^0(\bs^{**}-\bs^*),
(\bs^{**}+\bs^*))= }\\
\ds{
-2\sum_{i\in I_\e(\bs^*)}x^*_i+{1\over 2}(\ti{\bf J}^0(\bs^{**}-\bs^*),
(\bs^{**}-\bs^*))}\\
\ds{
\ge(1+2\alpha)\e^2 N-{\alpha\over 2}((\bs^{**}-\bs^*),(\bs^{**}-\bs^*)) }\\
\ds{
\ge(1+2\alpha)\e^2 N-2\alpha\e^2 N\ge\e^2 N},
\end{array}\label{T3.2}\end{equation}
where $\ti{\bf J}^0$ is defined by (\ref{tiJ}) with zero diagonal elements
 and we have used the inequality $\ti{\bf J}^0+\alpha {\bf I}=\ti{\bf J}\ge 0$.

So, we have proved that
\begin{equation}
{\cal A}\supset \cup_{\e>0}{\cal B_\e},
\label{T3.3}\end{equation}
where ${\cal B}_\e$ denotes the event, that for any extreme point
$\bs^*\in\Od^{1}$, the number $N_\e$ of indexes in the set  $
I_\e(\bs^*)$ is larger than $\e N$. Hence,
\begin{equation}
\ov{\cal A}\subset \cap_{\e>0}\ov{\cal B}_\e, \quad \P(\ov{\cal
A})\le\inf_{\e>0}\P(\ov{\cal B}_\e\cap{\cal K}_{\ti\e})+
\P\{\ov{\cal K}_{\ti\e}\}, 
\label{T3.4}\end{equation} 
where the event ${\cal K}_{\ti\e}$ means that  inequalities (\ref{ineqJ})
hold. Let us note now that $\ov{\cal B}_\e$ corresponds to the
event, that there exists a local minimal point $\bs^*\in\Od^{1}$, such
that $N_\e\le N\e$. Thus,
\begin{equation}
\P(\ov{\cal B}_\e\cap{\cal K}_{\ti\e})\le \sum_{k=0}^{[\e
N]}C_N^{[\delta N]} C_{[\delta N]}^k\P({\cal B}^0_{\e,k}\cap{\cal
K}_{\ti\e}), \label{T3.5}\end{equation} 
where ${\cal B}^0_{\e,k}$
denotes the event, that the point $\bs^{(1,\delta)}$ of the form
(\ref{x_k}) is a local minimal point in $\Od^{1}$, and $\ti x_i^0\le
-({1\over 2}+\alpha)\e$ for $i=1,\dots, k$. Taking into account
that under condition (\ref{ineqJ}) the necessary condition for
$\bs^{(1,\delta)}$ to be a minimum point
 is (\ref{condmin}), we obtain that for $k\not=0$,
\begin{equation}\begin{array}{c}\ds{
\P({\cal B}^0_{\e,k}\cap{\cal K}_{\ti\e})\le \P\{\ti x_i^0\ge -({1\over 2}+\alpha)\e,\,\,
i=k+1,\dots,[\delta N];}\\
\ds{
\ti x_j^0\ge -\ti\e,\,\,j=[\delta N]+1,\dots,N\}=
 P_{N,{k}}(-({1\over 2}+\alpha)\e,-\ti\e)}.
\end{array}\label{T3.6}\end{equation}
And for $k=0$,
\begin{equation}
{\cal B}^0_{\e,0}\cap{\cal K}_{\ti\e}\subset(\cap_{i=1}^{[\delta N]}
{\cal A}_i^0(-({1\over 2}+\alpha)\e)\cap_{j=[\delta N]+1}^N{\cal A}_j^0(-\ti\e))\cup
(\cup_{q>-\e(0.5+\alpha)}{\cal C}(\ti q)),
\label{T3.7}\end{equation}
where ${\cal A}^0_j(\ti q)$ is defined by (\ref{calA}) and
\begin{equation}
{\cal C}( q)\equiv\{\min_{i=1,\dots,[\delta N]}\ti x_i^0\ge q,
\min_{j=[\delta N]+1,\dots,N}\ti x_j^0=-q-\ti\e\}.
\label{T3.8}\end{equation}
But it is easy to see that for any $\Delta>0$, if we denote
$$
{\cal A}( q,- q-\Delta)\equiv \cap_{i=1}^{[\delta N]}
{\cal A}_i^0( q)\cap_{j=[\delta N]+1}^N{\cal A}_j^0(- q-\Delta-\ti\e),
$$
 then
\begin{equation}\begin{array}{c}\ds{
\cup_{0\le t\le 1}{\cal C}( q+t\Delta)\subset
{\cal A}( q,- q-\Delta-\ti\e)\Rightarrow }\\
\ds{
\P\{\cup_{0\le t\le 1}{\cal C}( q+t\Delta)\}\le P_N( q,
- q-\Delta-\ti\e). }
\end{array}\label{T3.9}\end{equation}
To have an upper bound for the value of $ q$ which we need to consider we
use
\begin{proposition}\label{pro:3}
For any positive $\a\le 0.113$  and $\de\le 0.6\a^2$ there exists $q_0(\a,\de)$, 
such that for any $\ti d>0$,
$$
\P\{\cup_{  q>q_0+\ti d}{\cal C}( q)\}\le\exp\{-NC_{\ti d}\},
$$
where $C_{\ti d}>C^*(\delta)$ with $C^*(\de)$  defined in (\ref{C^*}).

For $\a\le 0.113$, $\de\le 0.00645$ and $\de\le 0.6\a^2$
$\quad q_0(\a,\de)\le 0.13$.
\end{proposition}

On the basis of this proposition, we can  restrict ourselves
by $0\le q\le q_0+\ti d$ and, using (\ref{T3.7})-(\ref{T3.9}),
write
\begin{equation}\begin{array}{c}\ds{
\P\{{\cal B}\cap{\cal K}_{\ti\e}\}\le  P_N(-({1\over 2}+\alpha)\e,-\ti \e)+
\sum_{l=1}^MP_N(l\Delta,-\ti\e-(l+1)\Delta)}\\
\ds{
\le P_N(-({1\over 2}+\alpha)\e,\ti\e)+
M\max_{0\le q\le q_0+\ti d}
\ti P_N( q,- q-\Delta-\ti\e)}+e^{-NC_{\ti d}},
\end{array}\label{T3.11a}\end{equation}
where $M={q_0+\ti d+\e(0.5+\alpha]\over\Delta}$. Now,  using
Theorem \ref{thm:1}, we get from
(\ref{T3.4}), (\ref{T3.5}) and (\ref{T3.11a}),
\begin{equation}\begin{array}{c}\ds{
\P(\ov{\cal A}\cap{\cal K}_{\ti\e})\le \exp\{-NC_{\ti d}\}}\\
\ds{(M+1) C_N^{[\delta N]}C_{[\delta N]} ^{[\e N]}(\exp\{
N[C(\a,\delta,\ti\e,\e,\Delta)+o(1)]\}, }
\end{array}\label{T3.12}\end{equation}
where
$$\begin{array}{c}\ds{
C(\a,\delta,\ti\e,\e,\Delta)=\max[\max_{0\le\delta_1\le\e}
\max_U{\cal F}_1^D(U;\a,\delta,\delta_1,-({1\over 2}+\alpha)\e,-\ti\e)
-{\a\over 2}\log\a+{\a\over 2};}\\
\ds{
\max_U\min_V{\cal F}_0^D(U,V;\a,\delta,-({1\over 2}+\alpha)\e,-\ti\e)
-{\a\over 2}\log\a+{\a\over 2};\,\,\, }\\
\ds{
\max_{ q>\e(o.5+\alpha)}\max_U\min_V{\cal F}_0^D(U,V;\a,\delta, q,
- q-\Delta-\ti\e)-{\a\over 2}\log\a+{\a\over 2}].}
\end{array}$$
Since ${\cal F}_0^D$ and ${\cal F}_1^D$ are continuous with respect to
$q$, $q'$, $\delta_1$, we get for $\Delta,\e\to 0$,
\begin{equation}
\ds{
\P(\ov{\cal A}\cap{\cal K}_{\ti\e})\le \exp\{N[C(\a,\delta,\ti\e,\ti d)+o(1)]\}+\exp\{-N(C_{\ti d}-C^*(\delta))\},}
\label{T3.13}\end{equation}
where
\begin{equation}
C(\a,\delta,\ti d,\ti\e)=
\max_{0\le q\le q_0+\ti d}\max_U\min_V\{{\cal F}_0^D(U,V;\a,\delta, q,
 -q-\ti\e)-{\a\over 2}\log\a+{\a\over 2}+C^*(\delta)\},
\label{T3.131}\end{equation}
and therefore
\begin{equation}\begin{array}{c}
\ds{
\P(\ov{\cal A})\le \exp\{N[C(\a,\delta,\ti\e,\ti d)+o(1)]\}+\exp\{-N(C_{\ti d}
-C^*(\delta))\}}\\
\ds{+ \P\{\ov{\cal K}_{\ti\e}\}\le
\exp\{N[C(\a,\delta,\ti\e,\ti d)+o(1)]\}}\\
\ds{ +\exp\{-N(C_{\ti d}-C^*(\delta))\}+
\exp\{-\const N\ti\e^2\}.}
\end{array}\label{T3.14}\end{equation}
Since $(C_{\ti d}-C^*(\delta))> 0$ for all $\ti d>0$, we conclude, that
if for some $\delta>0$, $C(\a,\delta,0,0)<0$, then we always can choose $\ti d$ and $\ti\e$ small enough
to provide that all the exponents in the r.h.s. of (\ref{T3.14}) are negative.
Thus, we obtain the statement of Theorem \ref{thm:3}.

\begin{proposition}\label{pro:4}
Consider the functions
\begin{equation}\begin{array}{c}\ds{
\Phi(U,q,\a,\de)\equiv\min_V\{{\cal F}_0(U,V;\a,\delta, q, -q)
-{\a\over 2}\log\a+{\a\over 2}+C^*(\delta)\},}\\
\ds{
\Phi_0(q,\a,\de)\equiv\max_U \Phi(U,q,\a,\de)\equiv\Phi(U(q,\a,\de),q,\a,\de).}
\end{array}\label{T3.15}\end{equation}
If for some $0.071\le\alpha_1\le\alpha_2\le\a_c$, $0.0035\le\de\le\de_c=
 0.00778$,
\begin{equation}\begin{array}{c}\ds{
\Phi_0(0,\a_2,\de)< 0,\quad
{\d\Phi\over \d q}(U_2,0,\a_2,\de)<0,\quad
{\d\Phi\over \d \a}(U_1,0,\a_2,\de)>0,}
\end{array}\label{T3.18}\end{equation}
then $\Phi_0(q,\a,\de)< 0$ for any $\a_1\le\a\le\a_2$ and $0\le q\le q_0$.
Here $U_1=U(0,\a_1,\de)<U_2=U(q_0,\a_2,\de)$.

If also $\de\le k_c\a^2$ ($k_c\equiv{\de_c\over\a_c^2}$) and
\begin{equation}
\max_{U\le\sqrt\a}\min_V{\cal F}_0^D(U,V;\a,\delta)+C^*(\de)-
{\a\over 2}\log\a+{\a\over 2}<0,
\label{T3.17}\end{equation}
then $C(\a,\delta,0,0)$ defined by (\ref{T3.131}) is negative.
\end{proposition}

From (\ref{T:3.cond}) it is easy to see that  to find $\a_c$ and $\de_c$
we should study the field of parameters $\a$, $\de$ where $\Phi_0(0,\a,\de)<0$.
Let us fix for the moment $\a$ and study the behaviour of the function 
$\Phi_0(0,\a,\de)$ as a function of
$\de$. We find, that it is negative for $0\le\de\le\de_1(\a)$ and
$\de_2(\a)\le\de\le\de_3(\a)$. But for $0\le\de\le\de_1(\a)$ $C(\a,\delta,0,0)$
defined by (\ref{T3.131})
cannot be negative, because if it is so, then according to Theorem 
\ref{thm:3}, there exists a minimum point inside the ball
$B_{\de_1}^1$. But by the virtue of  Theorem 1,
the probability to have the minimum point in ${\Omega}^1_\de$ 
($\de<\de_1$) vanishes, as $N\to\infty$,
because $\Phi_0(0,\a,\de)<0$. Thus we should study 
$\de_2(\a)\le\de\le\de_3(\a)$.
When $\a$ increases,  $|\de_3(\a)-\de_2(\a)|$ decreases and  for
$\a=\a_c$ $\de_3(\a_c)=\de_2(\a_c)=\de_c$. Then evidently
$$
\Phi_0(0,\a_c,\de_c)=0,\quad {\d\Phi_0\over \d \de}(0,\a_c,\de_c)=0.
$$
So we find from these equations, that $\a_c=0.11326...$, $\de_c=0.00777...$
Unfortunately, for this $(\a_c,\de_c)$ condition (\ref{T3.17})
is  not fulfilled. So we take a bit smaller $\a=0.113$  and $\de=0.00645$,
for which (\ref{T3.17}) is fulfilled.  Then,
using (\ref{T3.18}), we  obtain  the statement of
Theorem 3 for all $0.071\le\a\le 0.113$ in three steps:

\no (1) $0.1105\le\a\le 0.113$, $\de=0.00645$;

\no (2)  $0.095\le\a\le 0.1105$, $\de=0.0042$;

\no (3) $0.071\le\a\le 0.095$, $\de=0.0035$.

 For $\a\le 0.071$
the statement of Theorem 3 follows from the result of \cite{Lou}.

\section{Auxiliary Results}

{\it Proof of Lemma \ref{lem:1}}.
At the first step we check that, if $\ti x_k$ are defined by
relations (\ref{x_k}), then
 $$
\la\theta(\ti x_k-(a_k+N^{1/2+d}))\ra\le e^{-\const N^{1+2d}}.
$$
To this end we use the Chebyshev inequality, according to which
$$\begin{array}{c}\ds{
\la\theta(\ti x_k-(a_k+N^{1/2+d})\ra\le\min_{\tau>0} \la \exp\{\tau
\ti x_k -\tau (a_k+N^{1/2+d})\}\ra}\\
\ds{
=\min_{\tau>0}e^{-\tau (a_k+N^{1/2+d})}
\prod_{\mu=1}^p\la\exp\{{\tau\over N}\sum_{j=1}^N\xmk\xmj\}\ra =
\min_{\tau>0}e^{-\tau (a_k+N^{1/2+d})}(\cosh{\tau\over
N})^{(pN)} }\\
\ds{
\le\min_{\tau>0}\exp\{-\tau
(a_k+N^{1/2+d})+\alpha{\tau^2\over 2}\}\le e^{-\const N^{1+2d}}.}
 \end{array}$$

Thus,
\begin{equation}\begin{array}{c}\ds{
\la\prod_{k=1}^N\theta(\ti x_k-a_k)\ra=
\la\prod_{k=1}^N\theta(\ti x_k-a_k)(\theta(a_k+N^{1/2+d}-\ti x_k) }\\
\ds{
+\theta(\ti x_k-(a_k+N^{1/2+d})))\ra\le
\la\prod_{k=1}^N\theta(\ti x_k-a_k)\theta(a_k+N^{1/2+d}-\ti
x_k)\ra }\\
\ds{
 2^N\sum_{k=1}^N\la\theta(\ti x_k-(a_k+N^{1/2+d}))\ra}\\
\ds{
\le\la\prod_{k=1}^N\theta(\ti x_k-a_k)\theta(a_k+N^{1/2+d}-\ti
x_k)\ra+ e^{-\const N^{1+2d}}. }
\end{array}\label{A0}\end{equation}

Consider
$$ \begin{array}{c}\ds{
D_{\l,\e_N^*}(x^1,...,x_N)\equiv}\\
\ds{
 {\exp\{-{1\over 2}\sum_{j,k=1}^N(\l{\bf I} +
\e_N^*l_N^{-1}{\bf J})^{-1}_{jk}x_j x_k-{1\over 2}\sum_{j,k=1}^N\e_N^*l_N^{-1}
J_{jk}\}\over l_N^{p/2}(2\pi)^{N/2}\hbox{det}^{1/2}\{\l{\bf I} +\e_N^*l_N^{-1}{\bf J}\}},}
 \end{array}$$
where ${\bf I}$ is a unit matrix and ${\bf J}$ is a matrix with
entries
$$
J_{jk}={1\over N}\sum_{\mu=1}^p\xmj\xmk.
$$
We study the composition $D_{\l,\e_N^*}*\prod\hN$ of this
function with the product of $\hN(x_k)$ (recall that $(f*g)(\ov x)\equiv\int f(\ov x-\ov x')
g(\ov x')d\ov x'$).
Let us check that for $0\le x_k\le N^{1/2+d}$,

\begin{equation}\begin{array}{c}\ds{
\prod_{k=1}^N\theta(x_k)\theta(N^{1/2+d}-x_k)\le
(1-e^{-h^2/2\l})^{-N}l_N^{p/2}}\\
\ds{
\cdot(D_{\l,\e_N^*}*\prod\hN)(x_1,...,x_N)\hbox{det}^{1/2}\{{\bf I} +
{\e_N^*\over\l l_N}{\bf J}\}\exp\{{\e_N^*\over 2l_N}
\sum_{j,k=1}^NJ_{jk}\}.}
\end{array}\label{A1}\end{equation}
Indeed, by definition of  composition,
 $$\begin{array}{c}\ds{
(D_{\l,\e_N^*}*\prod\hN)(x_1,...,x_N)\hbox{det}^{1/2}\{\l{\bf I} +
{\e_N^*\over l_N}{\bf J}\}\exp\{{\e_N^*\over 2l_N}
\sum_{j,k=1}^NJ_{jk}\}l_N^{p/2}}\\
\ds{
={1\over(2\pi)^{N/2}}\int\exp\{-{1\over 2}\sum_{j,k=1}^N(\l{\bf I}
+{\e_N^*\over l_N}{\bf J})^{-1}_{jk}(x_j-x_j')(
x_k-x_k')\}\prod_{k=1}^N\hN(x_k')dx_k'}
\end{array}$$
\begin{equation}\begin{array}{c}
\ds{
\ge{1\over(2\pi)^{N/2}}\int\exp\{-{1\over 2\l}\sum_{k=1}^N(x_k-x_k')^2\}
\prod_{k=1}^N\hN(x_k')dx_k'}\\
\ds{
\ge({1\over\sqrt{2\pi}}\int dx'
\exp\{-{(x-x')^2\over 2\l}\}\hN(x'))^N .}
\end{array}\label{A2}\end{equation}
But for $x\in (0,N^{1/2+d})$,
$$\begin{array}{c}\ds{
I_1=\int dx'
\exp\{-{(x-x')^2\over 2\l}\}(1-\hN(x'))}\\
\ds{
=\int_{-\infty}^{-h}
\exp\{-{(x-x')^2\over 2\l}\}dx'+
\int_{N^{1/2+d}+h}^{\infty}
\exp\{-{(x-x')^2\over 2\l}\}dx'}\\
\ds{
\le\int_{-\infty}^{-h}\exp\{-{(x')^2\over 2\l}\}dx'+
\int_{h}^{\infty}\exp\{-{(x')^2\over 2\l}\}dx'\le {2\l\over h}
e^{-{h^2\over 2\l}} .}
\end{array}$$
So for $h>({2\l\over \pi})^{1/2}$,
$$
{1\over\sqrt{2\pi}}\int dx'
\exp\{-{(x-x')^2\over 2\l}\}\hN(x')=(\sqrt\l-{I_1\over\sqrt
{2\pi}})\ge\sqrt\l(1-e^{-h^2/2\l}).
$$
Thus, we have proved (\ref{A1}) for $x_k\in(0,N^{1/2+d})$.
Besides, using the inequality $\log(1+x)\le x$, we get
\begin{equation}\begin{array}{c}\ds{
\hbox{det}^{1/2}\{{\bf I} +
{\e_N^*\over \l l_N}{\bf J}\}=
\exp\{{1\over 2}\sum_{\l_i\in\sigma({\bf J})}
\log(1+{\e_N^*\over \l l_N}\l_i\}}\\
\ds{
\le\exp\{{1\over 2}\sum_{\l_i\in\sigma({\bf
J})}{\e_N^*\over \l l_N}\l_i\}=\exp\{{\e_N^*\over 2\l l_N}\hbox{Tr}{\bf
J}\}=\exp\{{\e_N^*\alpha\over 2\l l_N}N\}.}
\end{array}\label{A3} \end{equation}
Here $\s({\bf J})$ is a spectrum of the matrix ${\bf J}$.

Therefore, it follows from (\ref{A1}) and (\ref{A3}) that for
$x_k\in(0,N^{1/2+d})$,
\begin{equation}\begin{array}{c}\ds{
\prod_{k=1}^N\theta(x_k)\theta(N^{1/2+d}-x_k)\le
(1-e^{-h^2/2\l})^{-N}l_N^{p/2}}\\
\ds{
\cdot\exp\{{\e_N^*\alpha N\over 2\l l_N}
\}(D_{\l,\e_N^*}*\prod\hN)(x_1,...,x_N) \exp\{{\e_N^*\over 2l_N}
 \sum_{j,k=1}^NJ_{jk}\}.}
\end{array}\label{A3.1}\end{equation}
But for all the
other values of $\{x_k\}$ the l.h.s. of this inequality is zero, while the
r.h.s. is positive, so we can extend (\ref{A3.1}) to all $\{x_k\}\in
{\bf R}^N$.

Besides, according to the  Chebyshev inequality,
\begin{equation}\begin{array}{c}\ds{
\P\{\sum J_{jk}\le
N(\e_N^*)^{-1/2}\}\le\min_{\tau>O}e^{-\tau(\e_N^*)^{-1/2}N}E\{e^{\tau\sum
J_{jk}}\}}\\
\ds{
=\min_{\tau>O}e^{-\tau(\e_N^*)^{-1/2}N}E^p\{\exp\{\tau\sum{1\over
N}\xi^1_j\xi^1_k\}\}}\\
\ds{
\le\min_{1>\tau>O}\exp\{-\tau
(\e_N^*)^{-1/2}N-{p\over 2}\log(1-\tau)\}}\\
\ds{
\le\exp\{-\const(\e_N^*)^{-1/2}N\}.}
\end{array}\label{A4}\end{equation}
Here we have used the standard trick, valid for $\tau<1$,
$$\begin{array}{c}\ds{
E\{\exp\{\tau\sum{1\over
N}\xi^1_j\xi^1_k\}\}=(2\pi)^{-1/2}E\{\int dx\exp\{-\sqrt\tau x
{1\over\sqrt N}\sum\xi^1_i-{x^2\over 2}\}}\\
\ds{
=(2\pi)^{-1/2}\int dx
(\cosh{x\sqrt\tau\over\sqrt
N})^Ne^{-x^2/2}=(1-\tau)^{-1/2}(1+O(N^{-1})).}
\end{array}
$$

Therefore finally, on the basis (\ref{A0}), (\ref{A3.1}) and
(\ref{A4}), we get
 \begin{equation}\begin{array}{c}\ds{
\la\prod_{k=1}^N\theta (\ti x_k-a_k)\ra\le
 e^{-N(\e_N^*)^{-1/2}\const}}\\
\ds{
+{e^{\const N(\e_N^*)^{1/2}}l_N^{p/2}\over (1-e^{-h^2/2\l})^{N}} \la
(D_{\l,\e_N^*}*\prod_{k=1}^N\hN)(\ti x_1-a_1,...,\ti
x_N-a_N)\ra.}
\end{array}\label{A5}\end{equation}

Now to finish the proof of Lemma \ref{lem:1} we are left to find the
Fourier transform $\hat D_{\l,\e_N^*}$ of the function  $ D_{\l,\e_N^*}$,
$$\begin{array}{c}\ds{
\hat D_{\l,\e_N^*}(\ov\zeta)=(2\pi)^{-N/2}\int d\ov x e^{i(\ov x,\ov\zeta)}
D_{\l,\e_N^*}(\ov x)=l_N^{-p/2}
\exp\{-{\l\over 2}(\ov\zeta,\ov\zeta)}\\
\ds{
-{\e_N^*\over
2l_N N}\sum_{\mu} (\sum_k\xmk\zk)^2-
{\e_N^*\over 2l_N N}\sum_{\mu} (\sum_k\xmk)^2\}}\\
\ds{
=l_N^{-p/2}\exp\{-{\l\over 2}(\ov\zeta,\ov\zeta)-{\e_N^*\over
2l_N }\sum_{\mu}((\ti \um)^2+(\ti \vm)^2)\},}
\end{array}$$
where $\ti \um$ and $\ti \vm$ are defined by (\ref{tiu}). Then
\begin{equation}\begin{array}{c}\ds{
\la (D_{\l,\e_N^*}*\prod_{k=1}^N\chi_{N,h})(\ti x_1-a_1,...,\ti x_N-a_N)\ra}\\
\ds{
=(2\pi)^{-N}\int \prod_{k=1}^N d\zk\hhN(\zk)\exp\{-ia_k\zk\}\cdot
\la\hat D_{\l,\e_N^*}(\ov\zeta)\exp\{i\sum_{k=1}^N\zk\ti x_k\}\ra}\\
\ds{
=l_N^{-p/2}(2\pi)^{-N}\int \prod_{k=1}^N d\zk\hhN(\zk)\exp\{-ia_k\zk-
{\l\over 2}\zk^2\}}\\
\ds{
\cdot\prod_\mu\la\exp\{-{\e_N^*\over 2l_N}(\ti \um)^2+(\ti \vm)^2+i\ti u\ti v
\}\ra.}
\end{array}\label{A5.1}\end{equation}
Let us use the representation (cf. (\ref{G2}) )
$$\begin{array}{c}\ds{
\la\exp\{-{\e_N^*\over 2l_N}((\ti \um)^2+(\ti \vm)^2)+i\ti\um\ti\vm\}\ra}\\
\ds{
={l_N^{1/2}\over 2\pi }\int d\um d\vm \la\exp\{-{\e_N^*\over 2}((\um)^2+
(\vm)^2)-il_N \um\vm+i\um\ti\um+i\vm\ti\vm\}\ra,}
\end{array}$$
where we have taken into account, that by definition (see Lemma \ref{lem:1})
$l_N=l_N^2+(\e_N^*)^2$. Substituting this representation
 into (\ref{A5.1}), we get
\begin{equation}\begin{array}{c}\ds{
\la (D_{\l,\e_N^*}*\prod_{k=1}^N\chi_{N,h})(\ti x_1-a_1,\dots,\ti x_N-a_N)\ra}\\
\ds{
=(2\pi)^{-N-p}\int \prod_{k=1}^N d\zk\hhN(\zk)\exp\{-{\l\over
2}\zk^2- ia_k\zk\}\prod_\mu\int d\um }\\
\ds{
\cdot d\vm\exp\{-i\um\vm-{\e_N^*\over 2}
(\um)^2-{\e_N^*\over 2}(\vm)^2\}\prod_{k=1}^N\cos{\um\zk+\vm\over
\sqrt N}}=P_N^1.
\end{array}\label{A5.2}\end{equation}
Inequality (\ref{A5}) and this representation prove Lemma \ref{lem:1}.

\medskip

 {\it Proof of Lemma \ref{lem:2}}.
 Take $L={\pi\over 6\e_N^2}$ and consider an intermediate functions:
\begin{equation}
\begin{array}{c}\ds{
F_{cL}^{(m)}(\uv_1,\vv_1,\uv_2,\vv_2)\equiv\int_{- L}^L
d\zk\hhN(\zk)e^{-\l\zk^2/2- ia\zk}}\\ \ds{
\cdot\prod_{\mu\le m}\cos
{\um\zk+\wm\over\sqrt N}\exp\{- {1\over
N}(\uv_2,\wv_2)\zk-{1\over 2N}(\uv_2,\uv_2)\zk^2\}\prod_{\nu>m}^p
\cos{\wn\over\sqrt N};}\\
\ds{
F_{NL}(\uv_1,\vv_1,\uv_2,\vv_2)\equiv\int_{- L}^L
d\zk\hhN(\zk)e^{-\l\zk^2/2- ia\zk}\prod_{\mu=1}^p\cos
{\um\zk+\wm\over\sqrt N}}.
\end{array}\label{A6}\end{equation}
Denote  also $F^{(m)}_c$  by the same formula as $F_{cL}^{(m)}$
 with $L=\infty$.

Then
\begin{equation}\begin{array}{c}\ds{
R^{(m)}\equiv F_N-F^{(m)}=(F_N-F_{NL})+(F_{NL}-F_{cL}^{(m)})}\\
\ds{
+(F_{cL}^{(m)}-F_{c}^{(m)})+(F_{c}^{(m)}-F^{(m)}).}
\end{array}\label{sumR}\end{equation}

One could easily estimate $(F_N-F_{NL})$
by using the simple inequalities
\begin{equation}
|(F_N-F_{NL})(\uv,\wv)|\le
{e^{{(\lv_2,\lv_2)\over N}}\over
2\pi}\int_{|\zk|>L}e^{-\zk^2/2\l}d\zk\le  e^{{(\lv,\lv)\over
N}}e^{-\const\e_N^{-4}}.
\label{A6.3}\end{equation}

 Let us estimate $R^{(m)}_*\equiv F_{NL}-F_{cL}^{(m)}$. To this end we consider
$$
f(\zk)=\sum_{\nu>m}\log\cos{\un\zk+\wn\over\sqrt N}+
{\zk^2\over 2} \ti U^2+{\zk\over N}(\uv_2,\wv_2)
$$
 and use the inequality
$$
|e^{f(\zk)}-e^{f(0)}|\le |f(\zk)-f(0)|(|e^{f(\zk)}|+|e^{f(0)}|).
$$
Then, since $|{\xi\un\over\sqrt N}|,|{\vn\over\sqrt N}|\le L\e_N^2
\le{\pi\over 6}$
and $|\un|,|\vn|,|\ln|\le\e_N\sqrt N$, we get
\begin{equation}\begin{array}{c}\ds{
|f(\zk)-f(0)|\le|\zk||f'(\xi)|}\\
\ds{
=|\zk||\sum_{\nu>m}[-{\un\over\sqrt
N}{\hbox{tg}}{\xi\un+\wn\over\sqrt N}+\xi{(\un)^2\over
N}+{\un\wn\over N}]|}\\
\ds{
\le|\zk|\const\sum_{\nu>m}|{\un\over\sqrt N}|
|{\xi\un+\wn\over\sqrt
N}|^3}\\
\ds{
\le\e_N^2|\zk|\const(\ti U^2|\zk|^3+{1\over
N}\sum_{\nu>m}(|\vn|^2+|\ln|^2))}.
\end{array}\label{A7}\end{equation}
To estimate $|e^{f(\zk)}|$ we use the inequality, valid for
$|\Re z|\le {\pi\over 2}$,
\begin{equation}
\Re(\log\cos z+{1\over 2}z^2)\le {1\over 2}(\Im z)^2.
\label{A8}\end{equation}
(The proof of this inequality is given at the end of the proof
 of Lemma \ref{lem:2}.)
It follows from (\ref{A8}) that
\begin{equation}\begin{array}{c}\ds{
\Re f(\zk)=\Re\{\sum_{nu=m+1}^p[\log\cos{\zk\un+\wn\over\sqrt N}+
{(\zk\un+\wn)^2\over 2N}-{(\wn)^2\over 2N}]\}}\\
\ds{
\le\sum_{\nu>m}{(\Im\{\zk\un+\wn\})^2\over 2N}-
\sum_{\nu>m}{\Re\{(\wn)^2\}\over 2N}=
 - {(\vv_2,\vv_2)\over
2N}+{(\lv_2,\lv_2)\over 2 N}}.
\end{array}\label{A9}\end{equation}

Therefore we derive from (\ref{A7}) and (\ref{A9}) that
\begin{equation}\begin{array}{c}\ds{
|\prod_{\nu>m}^p\cos
{\un\zk+\wn\over\sqrt N}-
\exp\{- {(\uv_2,\vv_2)\zk\over N}-{(\uv_2,\uv_2)\zk^2\over 2N}\}
\prod_{\nu>m}
\cos{\wn\over\sqrt N}|}\\
\ds{
=\exp\{- {1\over N}(\uv_2,\vv_2)\zk-{1\over 2N}(\uv_2,\uv_2)\zk^2\}
|e^{f(\zk)}-e^{f(0)}|}\\
\ds{
\le\const\e_N^2
|\zk|(\ti U^2|\zk|^3+{(\vv_2,\vv_2)+(\lv_2,\lv_2)\over N})}\\
\ds{
\cdot[\exp\{-{\zk^2\ti U^2\over 2}-\zk{(\uv_2,\vv_2)\over N}-
{(\vv_2,\vv_2)\over N}+{(\lv_2,\lv_2)\over N}\}}\\
\ds{
+\exp\{-{\zk^2\ti U^2\over 2}-\zk{(\uv_2,\vv_2)\over
N}\}\prod_{\nu>m}|\cos{\vn+i\ln\over\sqrt N}|].}
\end{array}\label{A10}\end{equation}
Using inequality (\ref{A8}) for $|\cos
{\vn+i\ln\over\sqrt N}|$ ($\nu>m$), we get
\begin{equation}\begin{array}{c}\ds{
|R^{(m)}_*(\uv_1,\vv_1,\uv_2,\vv_2+i\lv_2)|\le \e_N^2
\int d\zk(\ti U^2|\zk|^3+{(\vv_2,\vv_2)+(\lv_2,\lv_2)\over N})
e^{-\l\zk^2/2}}\\
\ds{
\cdot\prod_{\mu\le m}|\cos{\um\zk+\wm\over\sqrt
N}|\exp\{-{\zk^2\ti U^2\over 2}-\zk{(\uv_2,\vv_2)\over N}-
{(\vv_2,\vv_2)\over 2N}+{(\lv_2,\lv_2)\over N}\}}\\
\ds{
\le\e_N^2\const(1+{(\vv_2,\vv_2)+(\lv_2,\lv_2)\over N\sqrt{\ti U^2+\l}})
\exp\{-{(\vv_2,\vv_2)\over 2N}+{(\uv_2,\vv_2)^2\over
N^2(\ti U^2+\l)}+{(\lv,\lv)\over N}\}.}
\end{array}\label{A11}\end{equation}

Now to obtain the estimate of
the form (\ref{11}) we use  (\ref{13.2}) and the inequality
$$
{(\vv_2,\vv_2)\over 2N}\le {2(\ti U^2+\l)\over\l}\exp\{
{\l(\vv_2,\vv_2)\over 4N(\ti U^2+\l)}\}.$$
Combining them with (\ref{A11}), we get
\begin{equation}\begin{array}{c}\ds{
|R^{(m)}_*(\uv_1,\vv_1,\uv_2,\vv_2+i\lv_2)|}\\
\ds{
\le\e_N^2\const(\ti U^2+\l)^{1/2}(1+{(\lv_2,\lv_2)\over N})\exp\{-
{\l(\vv_2,\vv_2)\over 4N(\ti U^2+\l)}+{(\lv,\lv)\over N}\}.}
\end{array}\label{A11.1}\end{equation}
To estimate $(F_{cL}^{(m)}-F_{c}^{(m)})$ we use again the inequality
(\ref{A8}) for $|\cos{\vn+i\ln\over\sqrt N}|$ ($\nu>m$),
\begin{equation}\begin{array}{c}\ds{
|F_{cL}^{(m)}(\uv,\wv)-F_{c}^{(m)}(\uv,\wv)|\le e^{{(\lv,\lv)\over N}}
e^{-{(\vv_2,\vv_2)\over 2N}}}\\
\ds{
\cdot\int_{|\zk|\ge L}
d\zk|\hhN(\zk)|e^{-\l\zk^2/2}
\exp\{- {1\over
N}(\uv_2,\vv_2)\zk-{1\over 2N}(\uv_2,\uv_2)\zk^2\}}\\
\ds{
\cdot \le e^{{(\lv,\lv)\over N}}\int_{|\zk|\ge L}
d\zk e^{-\l\zk^2/2}\le\const e^{{(\lv,\lv)\over N}}e^{-\const\e_N^{-4}}}.
\end{array}\label{A11a}\end{equation}
Thus, we are left to estimate the difference
\begin{equation}\begin{array}{c}\ds{
F^{(m)}_c(\uv_1,\wv_1,\uv_2,\wv_2)-F^{(m)}(\uv_1,\wv_1,\uv_2,\wv_2)
}\\
\ds{=\la
H_{N,h,\ti U}({a-i(\uv_2,\wv_2)-
{(\uv_1,\ov\xi_1)\over\sqrt N}\over\sqrt{\ti U^2+\l}})
e^{i(\vv_1,\ov\xi_1)\over\sqrt N}\ra
(\prod_{\mu>m}\cos{\wn\over\sqrt N}-e^{-{(\wv_2,\wv_2)\over2N}}).}
\end{array}\label{A14}\end{equation}

The last multiplier here can be estimated by the same way as
in (\ref{A6})-(\ref{A10}). Then we get
\begin{equation}\begin{array}{c}\ds{
|\prod_{\mu>m}\cos{\wn\over\sqrt N}-e^{-{(\wv_2,\wv_2)\over2N}}|}\\
\ds{
\le\const\e_N^2{|(\wv_2,\wv_2)|\over N}\exp\{-{(\vv_2,\vv_2)\over 2N}+
{(\lv_2,\lv_2)\over N}\}.}
\end{array}\label{A15}\end{equation}
To estimate the first multiplier we use the bound
$|H_{N,h,\ti U}(a+ic)|\le e^{c^2/2}$. Thus,
$$\begin{array}{c}\ds{
\la
H_{N,h,\ti U}({a-i(\uv_2,\wv_2)-{(\uv,\ov\xi_1)\over\sqrt N}
\over\sqrt{\ti U^2+\l}})e^{{i(\vv_1,\ov\xi_1)\over\sqrt N}}\ra}\\
\ds{
\le\exp\{{(\uv_2,\vv_2)^2\over
2N^2(\ti U^2+\l)}\}\la e^{{(\lv_1,\ov\xi_1)\over\sqrt N}}\ra\le
\exp\{{(\lv_1,\lv_1)\over N}+{(\uv_2,\vv_2)^2\over
2N^2(\ti U^2+\l)}\}.}
\end{array}$$
By the same way as in (\ref{A10})-(\ref{A11.1}) we
can obtain now from (\ref{A14}) and (\ref{A15}) the bound of the
form (\ref{11}).

Now to finish the proof of Lemma \ref{lem:2} we are left to prove
inequality (\ref{A8}). For $z=x+iy$ ($x,y\in{\bf R}$) by the simple algebraic
transformations we get that (\ref{A8}) is equivalent to the inequality
\begin{equation}
{1\over 2}(\cosh 2y+\cos 2x)\le e^{2y^2-x^2}.
\label{A8.1}\end{equation}
Since $\cosh 2y\le e^{2y^2}$, to prove (\ref{A8.1}) it is enough to prove that
$$
\cos 2x\le e^{2y^2}(2 e^{-x^2}-1),
$$
which evidently follows from
$$
\cos 2x\le (2 e^{-x^2}-1)\quad\iff\quad \cos x\le e^{-x^2/2}.
$$
Since the last inequality is valid for $|x|\le{\pi\over 2}$, we have proved
(\ref{A8.1}) and so (\ref{A8}).

Lemma \ref{lem:2} is proven.

\medskip

\no{\it Proof of Lemma \ref{lem:3}.}
 We use  (\ref{11}) to estimate the integral
$$\begin{array}{c}\ds{
I'_{m,k}\equiv\int_{-\e_N\sqrt
N}^{\e_N\sqrt N}d\vv_2e^{-il_N(\uv_2,\vv_2)}e^{-{\e_N^*\over
2}(\vv_2,\vv_2)}}\\
\ds{
\cdot(F^{(m)}(\uv_1,\vv_1,\uv_2,\vv_2))^{N-k}
(R^{(m)}(\uv_1,\vv_1,\uv_2,\vv_2))^{k}. }
\end{array}$$

By using (\ref{boundH}), which is evidently valid also for $H_{N,h,\ti U}$
we get
\begin{equation}\begin{array}{c}\ds{
|F^{(m)}(\uv_1,\vv_1,\uv_2,\vv_2)|\le
\exp\{{(\uv_2,\vv_2)^2\over 2N^2(\ti U^2+\l)}-{(\vv_2,\vv_2)\over 2N}\}}\\
\ds{
\le \exp\{-{\l(\vv_2,\vv_2)\over 2N(\ti U^2+\l)}\}\le
\exp\{-{\l(\vv_2,\vv_2)\over 4N(\ti U^2+\l)}\}. }
\end{array}\label{13.2}\end{equation}
The second inequality here can be obtained if we observe that
${(\uv_2,\vv_2)^2\over N^2(\ti U^2+\l)}=$
${\ti U^2\over \ti U^2+\l}({\bf P}_u\vv_2,\vv_2)$,
 where ${\bf P}_u$ is the orthogonal
projection operator on the unit vector $(\ti U)^{-1}N^{-1/2}\uv_2$, and use
the trivial inequality ${\bf I}-{\ti U^2\over \ti U^2+\l}{\bf P}_u\ge$
${\l\over \ti U^2+\l}{\bf I}$. Note also, that
we replace in (\ref{13.2}) 2 in the denominator by 4 in order to
have the same factor as in (\ref{11}).
Hence, on the basis of Lemma \ref{lem:2}, we have
\begin{equation}\begin{array}{c}\ds{
|I'_{m,k}|\le
\int_{-\e_N\sqrt N}^{\e_N\sqrt N}d\vv_2
|(F^{(m)}(\uv_1,\vv_1,\uv_2,\vv_2))^{N-k}
(R^{(m)}(\uv_1,\vv_1,\uv_2,\vv_2))^{k}|}\\
\ds{
\le e^{k\const}\e_N^{2k}(\ti U^2+\l)^{k/2} \int_{-\e_N\sqrt
N}^{\e_N\sqrt N}d\vv_2
\exp\{-{\l(\vv_2,\vv_2)\over 4(\ti U^2+\l)}-\e_N^*{(\vv_2,\vv_2)\over 2}\}}\\
\ds{
+e^{k\const}e^{-k\const\e_N^{-4}} \int d\vv_2
\exp\{-{\l(N-k)(\vv_2,\vv_2)\over 4N(\ti U^2+\l)}-
\e_N^*{(\vv_2,\vv_2)\over 2}\} }\\
\ds{
\le e^{N\const}(\ti U^2+\l)^{p/2}\e_N^{2k}+e^{k\const}(\e_N^*)^{-p/2}
e^{-k\const\e_N^{-4}} .}
\end{array}\label{13}\end{equation}
Substituting estimate (\ref{13}) in the expression for $I_{m,k}$
integrating over $\uv_1,\vv_1$, and $\ti U$
we get finally
$$\begin{array}{c}\ds{
|I_{m,k}|\le
(\int  (\ti U^2+\l)^{p/2}\ti U^{p-m} e^{-N\e_N^*U^2/2}d\ti U)
e^{N\const}(\e_N)^{2k}}\\
\ds{
+e^{k\const}(\e_N^*)^{-p}
e^{-k\const\e_N^{-4}}.}
\end{array} $$
Using the Laplace method for the integration with respect to
$\ti U$ and taking into account that the second term in the r.h.s.
here  for $k>k_0$ is much smaller than the first one, we obtain
the statement of Lemma \ref{lem:3}.

\medskip

\no{\it Proof of Lemma \ref{lem:4}.}
 To prove (\ref{17}) we use the  variables
$\wm=-i\lm+\tm$, $(\tm\in{\bf R})$ $(\mu=1,...,m)$ and
$\wn=- i{\un\over \ti U}V+\tn$, $(\tn\in{\bf R})$
$(\nu=m_0+1,...,p-n)$ defined in (\ref{14}) and estimate
$$\begin{array}{c}\ds{
|G_{m,k,n}(V,\uv_1,\lv_1,\uv_3)| }\\
\ds{
\le\sum_{k_1+k_2=k}C_k^{k_1} (2\pi)^{-p}\int_{-\e_N^2\sqrt
N}^{\e_N^2\sqrt N} d\uv_4\int_{\prod L^\nu_2}|d\wv_4 |
\int d\tv_1\int_{-\e_N\sqrt N}^{\e_N\sqrt N}d\tv_3}\\
\ds{
|\la e^{{(\lv_1,\ov\xi_1)\over\sqrt N}}
H_{N,h,\ti U}({a_1-V{U^2\over\ti U}-i{(\uv_3,\tv_3)\over N}-
i{(\uv_4,\wv_4)\over N}
-{(\uv_1,\ov\xi_1)\over\sqrt N}\over\sqrt{U^2+\l+
N^{-1}(\uv_4,\uv_4)}}
\ra|^{[N\delta]-k_1} }\end{array}$$
\begin{equation}\begin{array}{c}\ds{
\cdot|\la e^{{(\lv_1,\ov\xi_1)\over\sqrt N}}
H_{N,h,\ti U}({a_2-V{U^2\over\ti U}-i{(\uv_3,\tv_3)\over N}-
i{(\uv_4,\wv_4)\over N}
-{(\uv_1,\ov\xi_1)\over\sqrt N}\over\sqrt{U^2+\l
+N^{-1}(\uv_4,\uv_4)}}
 )\ra|^{N-[N\delta]-k_2} }\\
\ds{
\cdot |R_{m}(\uv,\wv)|^{k}
\exp\{-l_N((\uv_1,\lv_1)+NV\ti U-\Im(\uv_4,\wv_4))}\\
\ds{
-(N-k)({1\over 2N}(\tv_3,\tv_3)
-V^2{U^2\over 2\ti U^2}+{1\over 2N}\Re(\wv_4,\wv_4))
 }\\
\ds{
-{\e_N^*\over 2}((\uv,\uv)+
(\tv_1,\tv_1)-(\lv_1,\lv_1)+(\tv_3,\tv_3)
- N V^2{U^2\over\ti U^2}+\Re(\wv_4,\wv_4))\}  .}
\end{array}\label{18}\end{equation}
Here we consider $I_{m,k}$
as the sum of terms, in which $k_1$ remainder functions $R^{(m)}$
come from the first $[\delta N]$ factors in (\ref{12}) and
$k_2$ of $R^{(m)}$ come from the last $N-[\delta N]$ ones.
Since $k=o(N)$ we have that $k_{1,2}=o(N)$ and $C_k^{k_1}=e^{o(N)}$.

Now we use (\ref{boundH}) for $H_{N,h,\ti U}$ and
the inequalities
\begin{equation}\begin{array}{c}\ds{
|N^{-1}(\uv_4,\wv_4)|\le N^{-1}\e_N\sqrt N\sum_{\nu=p-n+1}^p|\un|+
{V\over N\ti U}\sum_{\nu=p-n+1}^p|\un|^2}\\
\ds{
\le n\e_N^3+n{V\over \ti U}\e_N^4\le\const\,   \e_N^{1/2};}\\
\ds{
0\le N^{-1}(\uv_4,\uv_4)= \ti U^2-U^2\le n\e_N^4\le
\e_N^{3/2},}
\end{array}\label{18.1}\end{equation}
which are valid
since $n\le\e_N^{-5/2}$, $|\un|\le \e_N^2\sqrt N$ (see formula (\ref{1.1}))
 and $|w^\nu|<\e_N\sqrt N+{V\over \ti U}|\un|$ ($\nu=p-n+1,\dots,p$).
Besides,  $\exp\{{\e_N^*\over
2}[(\lv_1,\lv_1)+NV^2{U^2\over \ti U^2}]\}\le$  $e^{N\const \e_N^*}=e^{o(N)}$
because of the chosen bounds on $\lv_1$ and $V$.
Then,  using the inequality
\begin{equation}
H_{N,h,\ti U}(x)\le H(x),
\label{H_N<H}\end{equation}
and the fact that $k_{1,2}=o(N)$, we get from (\ref{18}),
\begin{equation}\begin{array}{c}\ds{
|G_{m,k,n}(V,\uv_1,\lv_1,\uv_3)|\le
{(\const \e_N^{2}\sqrt{U^2+\l})^{k}\over(2\pi)^{p}}
e^{-nN\e_N^2/4}}\\
\ds{
\cdot( G_m^*(U,V,\uv_1,\lv_1))^N
\exp\{-{\e_N^*\over 2}(\uv_1,\uv_1)-N{\e_N^*\over 2}U^2+No(1)\}}\\
\ds{
\cdot\int d\tv_1
d\tv_3\int_{-\e_N^2\sqrt N}^{\e_N^2\sqrt N}
d\uv_4
\exp\{-{N-k-n\over
2N}[(\tv_3,\tv_3)-{(\uv_3,\tv_3)^2\over N(U^2+\l)}]}\\
\ds{
-{\e_N^*\over 2}
((\tv_1,\tv_1)+(\tv_3,\tv_3)+(\uv_4,\uv_4))\}.}
\end{array}\label{19}\end{equation}
Here the term $(\const \e_N^2\sqrt{U^2+\l})^k$ is due to Lemma \ref{lem:2}
and the last line of (\ref{18.1}), and the term $e^{-nN\e_N^2/4}$ is due
to the integration with respect to $\wv_4$. On the other hand, we
should note that in fact   integrals with respect $\tv_1$ and
$\uv_4$ can give us only $(\const)^{m+n}(\e_N^*)^{-(m+n)}$ as a
multiplier. Since $m,n=o(N |\log \e_N^*|^{-1})$, we take it into
account as $e^{o(N)}$. Our main problem is to estimate the
integral with respect $\tv_3$, because it contains almost $p$
integrations. To perform this integration let us note that it is
of the Gaussian type  with the matrix of the form
${\bf A}=({\bf I}-{\ti U^2\over \ti U^2+\l}{\bf P}_u)$, where
${\bf I}$ is a unit matrix and ${\bf P}_u$ is the orthogonal
projector on the normalized vector ${\uv_3\over\sqrt{N}U}$. Since
such a matrix ${\bf A}$ has $(p-m-n-1)$ eigenvalues equal to 1 and
only one eigenvalue equal to $1-{\ti U ^2\over \ti U^2+\l}=
{\l\over\ti U^2+\l}$, the integration with respect to $\tv_3$
gives us $(2\pi)^{p-n-m\over2}\const$. Thus we obtain (\ref{17}).

\no{\it Proof of Proposition \ref{pro:1}.}
 It follows from (\ref{28a}) that
\begin{equation}\begin{array}{c}\ds{
\log |G_{m}^*(U,V,\uv_1,\lv_1(U,V,\uv_1))|}\\
\ds{
\le N[o(1)-UV+{1\over 2}V^2]
+C(U,V)-D(U,V)(\uv_1,\uv_1),}
\end{array}\label{P1}\end{equation}
where
$$
C(U,V)=N \delta \log
H({a_1-h-UV\over  \sqrt{U^2+\l}})+ N(1-\delta)
\log H({a_2-h-UV\over \sqrt{U^2+\l}}).
$$
On the other hand, using that $ H(x)<1$, we get
$$\begin{array}{c}\ds{
\la H({a_{1,2}-h-VU-
{(\uv_1,\ov\xi_1)\over\sqrt N}\over\sqrt{U^2+\l}})
e^{{(\lv_1,\ov\xi_1)\over\sqrt N}}\ra
\le\la e^{{(\lv_1,\ov\xi_1)\over\sqrt N}}
\ra\le e^{{(\lv_1,\lv_1)\over 2 N}}. }
\end{array}$$
Therefore, taking in (\ref{16}) $\lm=\um$  we obtain
\begin{equation}\begin{array}{c}\ds{
\log |G_{m}^*(U,\uv_1,\lv_1(U,V,\uv_1))|\le
N[-UV+{1\over 2}V^2]-{1\over 2}(\uv_1,\uv_1).}
\end{array}\label{P2}\end{equation}
Inequalities (\ref{P1}) and (\ref{P2}) give us
\begin{equation}\begin{array}{c}\ds{
\log |G_{m}^*(U,V,\uv_1,\lv_1(U,V,\uv_1))|\le
N(o(1)-UV+{1\over 2}V^2)}\\
\ds{
+\min[C(U,V)- D(U,V)(\uv_1,\uv_1);-{1\over 2 }(\uv_1,\uv_1)].}
\end{array}\label{P3}\end{equation}
Now, applying the Laplace method, we get
\begin{equation}\begin{array}{c}\ds{
\int d\uv_1 |G_{m}^*(U,V,\uv_1,\lv_1(U,V,\uv_1))| \exp\{
-{\e_N^*N\over 2}\sum_{\mu=1}^m(\um)^2\}}\\
\ds{
 \le\exp\{N(-UV+{1\over 2}V^2+o(1))}\\
\ds{
+\max_{(\uv_1,\uv_1)}\min[ C(U,V)-D(U,V)(\uv_1,\uv_1);
\quad -{1\over 2 }(\uv_1,\uv_1)]\}.}
\end{array}\label{P4}\end{equation}
But since both functions in the r.h.s. of (\ref{P4}) are linear ones
with respect
to $(\uv_1,\uv_1)$, one can find the maximum value explicitly.
It is just the intersection point of two functions
$y=-{1\over 2}x$ and $y=C(U,V)-D(U,V)x$.
It is easy to see that
$$
x_{int}=-{C(U,V)\over 0.5-D(U,V)},\quad
y_{int}={C(U,V)\over 1-2D(U,V)}.
$$
Substituting $y_{int}$ in (\ref{P4}) we get the statement of
Proposition 1.

\bigskip

\no {\it Proof of Proposition \ref{pro:2}.}
 The inequality $V(U)<U$ follows easily from (\ref{eqV}),
if we take into account, that $A(x)>0$. To prove that
$V(U)\ge \sqrt\alpha$ we use the inequalities:
\begin{equation}
0<A'(x)<1,\quad A(x+y)<A(x)+y<1+y\,\ (x<0,\,\,y>0).
\label{ineqA}\end{equation}
From the relations
$$\begin{array}{c}\ds{
A'(x)={e^{-x^2/2}\over 2\pi H^2(x)}(e^{-x^2/2}-x\sqrt{2\pi}H(x)),\quad}\\
\ds{
\sqrt{2\pi}H(x)x\le\int_x^\infty te^{-t^2/2}dt=e^{-x^2/2},}
\end{array}$$
it is easy to derive that $A'(x)>0$.
To get the upper bound for $A'(x)$ let us introduce  the function
$\phi(x)\equiv\log H(x)+{x^2\over 2}$. Using the identities
$$
\phi(x)=\log\int_0^\infty{dt\over\sqrt {2\pi}}e^{-tx-t^2/2},\quad
\phi''(x)=\la(t-\la t\ra_x)^2\ra_x\ge 0,
$$
where $\ds{\la...\ra_x\equiv{\int_0^\infty(...)e^{-tx-t^2/2}dt\over
\int_0^\infty e^{-tx-t^2/2}dt}}$, we obtain that $A'(x)\equiv 1-\phi''(x)<1$.

The last bound in (\ref{ineqA}) can be obtained as
$$
A(x+y)\le A(x)+y\max_{x\le s\le x+y}|A'(s)|<A(x)+y.
$$
Taking into account, that  $A(x)<\sqrt{2\over\pi}<1$ for $x<0$, we
get the last inequality in (\ref{ineqA}).

 Now from the bound $A'(x)< 1$ we get that the r.h.s. of
(\ref{eqV}) is an increasing function with respect to $V$.
Thus, to prove Proposition 2 it is enough to prove that
\begin{equation}
U>\sqrt\alpha+\delta A({\alpha+p\over U}-\sqrt\alpha)+
(1-\delta)A({\alpha-p\over U}-\sqrt\alpha),
\label{U>}\end{equation}
 for $U\ge 2\sqrt{\alpha}$. Here and below we denote
$p=1-2\delta$.

Using the last inequality in (\ref{ineqA}) with $x=-\sqrt\alpha$ and
$y={\alpha+p\over U}$
to estimate the first $A$, we get
\begin{equation}\begin{array}{c}\ds{
\delta A({\alpha+p\over U}-\sqrt\alpha)+(1-\delta)A({\alpha-p\over U}-\sqrt\alpha)<
\delta({\alpha+p\over U}+1)}\\
\ds{
+0.3{U(1-\delta)\over p-\alpha+2\alpha }<\delta
({\alpha+p\over 2\sqrt\alpha}+1)+
{0.3U\over 1+O(\alpha) }}\\
\ds{
=0.3U(1+O(\alpha))+o(\alpha^2)}.
\end{array}\label{rhs}\end{equation}
Here in order to estimate the second $A$ in (\ref{U>}) we have used
the bound $\max_xxA(-x)<0.3$,
which can be easily checked numerically.
It implies
$$
A(-{p-\alpha+\sqrt\alpha U\over U})<0.3
{U\over p-\alpha+2\alpha}\le
{0.3U\over 1+O(\alpha) }.
$$
So, if $U>2\sqrt{\alpha}$, then
\begin{equation}
U>\sqrt\alpha+0.3U(1+O(\alpha))+o(\alpha^2),
\label{condq1}\end{equation}
and (\ref{U>}) is valid. Thus, we have finished
the proof of Proposition \ref{pro:2}.

\bigskip

\no {\it Proof of Proposition \ref{pro:3}.}
Since for any $\ti q>q$ ${\cal C}(\ti q)\subset\cap_{j=1}^{[\de N]}\{\ti x^0_j
\ge q\}$, on the basis of Theorem 1, we have got
\begin{equation}\begin{array}{c}\ds{
\P\{\cup_{ \ti q>q}{\cal C}(\ti q)\}}\\
\ds{
\le\exp\{ N \max_{U>0}\min_{V}{\cal F}_0^D(U,V;\a,\delta,q,-\infty)
-{\alpha\over 2}\log\alpha+
{\alpha\over 2}\}.}
\end{array}\label{pro:3.1}\end{equation}
Let us denote 
$$\begin{array}{c}\ds{
f_0(U,V;q,\a,\de)\equiv{\cal F}_0(U,V;\a,\delta,q,-\infty)
+{\a\over 2}\log\a+{\a\over 2}+C^*(\de)}\\
\ds{
f^D(U,V;q,\a,\de)\equiv {\a\over 2}\log\a+{\a\over 2}+C^*(\de)
+\a\log U-UV}\\
\ds{+{V^2\over 2}+\de{\log H(a_1^*U^{-1}-V)\over 1-2D(U,V)},}
\end{array}$$
 and  consider
\begin{equation}\begin{array}{c}\ds{
\max_U\min_V f_0(U,V;q,\a,\de)\le
\max_Uf_0(U,U;q,\a,\de)}\\
\ds{
\le\max_U\{\a\log U-{U^2/2}-{\de\over 2}({a_1^*\over U}-U)^2\}+
{\a\over 2}\log\a+{\a\over 2}+C^*(\de)\to -\infty,}
\end{array}\label{pro:3.2}\end{equation}
as $a_1^*\to\infty$.
Here we have used the inequality $\ds{\log H(x)\le-{x^2\over 2}}$ ($x>0$).
 Similarly, for $f^D(U,V;q,\a,\de)$, when
$D(U,V)<0$ we have the bound
\begin{equation}\begin{array}{c}\ds{
\max_U\min_V f^D(U,V;q,\a,\de)\le
\max_U f^D(U,U;q,\a,\de)}\\
\ds{
\le\max_U\{\a\log U-{U^2/2}-{U^2\over 2}{A(a_1^*U^{-1}-U)\over
2U+(1-\de)A(a_1^*U^{-1}-U)}\} }\\
\ds{
-{\a\over 2}\log\a+{\a\over 2}+C^*(\de)
\le\max_U\{\a\log U-{U^2/2}}\\
\ds{
-{U^2\over 2}{p-U^2\over p(1-\de)+U^2(1+\de)}\}
-{\a\over 2}\log\a+{\a\over 2}+C^*(\de)
\to -{\a\over 2}\log 2+C^*(\de).}
\end{array}\label{pro:3.3}\end{equation}
Here we have used the inequalities $\ds{\log H(x)\le -{A(x)^2/2}}$
($x>0$) and $A(x)\ge x$.
Thus, inequalities (\ref{pro:3.2}) and (\ref{pro:3.3}) under conditions
$\de\le 0.6\a^2$, $\a\le 0.113$ prove the first statement of Proposition \ref{pro:3}.
Besides, (\ref{pro:3.3}) shows that it is enough to study only $f_0$.
Since $\max_U\min_Vf_0(U,V;q,\a,\de)$ for fixed $p$
increases with $\a$ and $\de$, to prove the second statement
of Proposition \ref{pro:3} it is enough to check that for
$\a=0.113$, $\de=\de_{max}=0.00645$ and $q=q_0+2\de_{min}-2\de_{max}=0.126$
$\max_U\min_Vf_0(U,V;q,\a,\de)<0$.
We do this numerically. Thus,
we obtain the statement of Proposition \ref{pro:3}.

\medskip

\no {\it Proof of Proposition \ref{pro:4}.}
 Let $I=I_U\times
I_\a\times I_q \subset{\bf R}^3$ with $I_U=[U_1,U_2]$, $I_\a=[\a_1,\a_2]$ and
$I_q=[0,q_0]$.
 Denote by $V(U,q,\a)$ the point of minimum of $\F(U,V;\a,\de,q,-q)$
 and by $U(q,\a)$ the point of maximum of $\Phi(U,q,\a)$. Let us note that
during the proof of Proposition \ref{pro:4} the variable $\de$ is fixed. So
here and below we omit $\de$ as an argument of the  functions $\Phi$ and
$\Phi_0$.

The first statement follows from the relations:
\begin{equation}\begin{array}{c}\ds{
U(q,\a)\in I_U \qquad (q\in I_q,\,\,\a\in I_\a),}\\
\ds{
 \Phi(U,q,\a)\le \Phi(U,0,\a)\le \Phi(U,0,\a_2)\le
\Phi(U(0,\a_2),0,\a_2)\le 0.}
\end{array}\label{P:4.1}\end{equation}
To prove the first line of (\ref{P:4.1}) it is enough to check that in $I$
\begin{equation}
{\d^2\Phi\over\d U\d\a}\ge 0,\,\,{\d^2\Phi\over\d U\d q}\ge 0,\quad
( 0\le q\le q_0,\,0.071\le\a\le 0.113),
\label{P:4.2}\end{equation}
because in this case we have for any $q\in I_q,\,\a\in I_\a$,
$$\begin{array}{c}\ds{
0={\d\Phi\over\d U}(U_1,0,\a_1))<{\d\Phi\over\d U}(U_1,q,\a_1)
<{\d\Phi\over\d U}(U_1,q,\a),}\\
\ds{
0={\d\Phi\over\d U}(U_2,q_0,\a_2)>{\d\Phi\over\d U}(U_2,q,\a_2)
>{\d\Phi\over\d U}(U_2,q,\a),}
\end{array}$$
and thus $U_1\equiv U(0,\a_1)\le U(q,\a)\le U(q_0,\a_2)\equiv U_2$.
Note, that for our choice of $0.0035\le\de\le 0.00778$,
$0.71\le\a\le 0.1133$ and $0\le q\le q_0\le 0.13$
 we get, that $0.25<U_1<U_2<0.41$.

Let us prove  (\ref{P:4.2}). To this end  we write
\begin{equation}\begin{array}{c}\ds{
{\d^2\Phi\over\d U\d \a}={\d^2\ti\F\over\d U\d\a}+
V'_\a{\d^2\ti\F\over\d U\d V};}\\
\ds{
{\d^2\Phi\over\d U\d q}={\d^2\ti\F\over\d U\d q}+
V'_q{\d^2\ti\F\over\d U\d V},}
\end{array}\label{P:4.4}\end{equation}
where
$\ti\F(U,V;\a,\de,q)\equiv\F(U,V;\a,\de,q,-q)-{\a\over 2}\log\a+{\a\over 2}$
and   $V'_{q,\a}$ are the derivatives with respect to $q$ and
$\alpha$ of the function $V(U,q,\a)$ defined above.
 By the standard method, from the equation
  $\ds{{\d\ti\F\over\d V}(U,V(q,\a))=0}$ we get
\begin{equation}
V'_\a=-({\d^2\ti\F\over\d V^2})^{-1}{\d^2\ti\F\over\d V\d\a},\quad
V'_q=-({\d^2\ti\F\over\d V^2})^{-1}{\d^2\ti\F\over\d V\d q}.
\label{P:4.5}\end{equation}
Now let us find the expressions for the derivatives of the function $\ti\F$,

$$\begin{array}{l}\ds{
{\d^2\ti\F\over\d V^2}=1-\de U^2A_1'-(1-\de)U^2A_2'>0;
\quad {\d^2\ti\F\over\d  q^2}=-\de A_1'-(1-\de)A_2'<0;}\\
\ds{
{\d^2\ti\F\over\d  \a^2}=-{1\over 2\a}-\de A_1'-(1-\de)A_2'<0;
\quad {\d^2\ti\F\over\d V\d \a}=\de UA_1'+(1-\de)UA_2'>0;\,\,}\\
\ds{
{\d^2\ti\F\over\d U\d \a}={1\over U}+{\de\over U}A_1+
{(1-\de)\over U}A_2+\de a_1^* A_1'+(1-\de)a_2^*A_2';}
\end{array}$$
\begin{equation}\begin{array}{c}
\ds{
{\d^2\ti\F\over\d U\d q}={\de\over U}A_1-
{(1-\de)\over U}A_2+{\de\over U}a_1^*A_1'-
{(1-\de)\over U}a_2^*A_2';}\\
\ds{
{\d^2\ti\F\over\d U\d V}=-1-\de a_1^*A_1'-
(1-\de)a_2^*A_2';}\\
\ds{
{\d^2\ti\F\over\d V\d q}=\de UA_1'-
(1-\de)UA_2';\,\,}
\end{array}\label{P:4.7}\end{equation}
where  $A_{1,2}$ are defined in (\ref{D(U)}) and
$$
A_{1,2}'\equiv{1\over U^2}A'({a_{1,2}^*\over U}-V)=
A_{1,2}(A_{1,2}-{a_{1,2}^*\over U^2}+{V\over U}),
$$
with  function $A(x)$ defined by (\ref{A(x)}). We recall here, that
from definition (\ref{a}), it follows that
\begin{equation}
1<a_1^*<1.25,\qquad -1.1<a_2^*<-0.85.
\label{bounda}\end{equation}
 Let us note also, that for $U\le U_2<0.41$,
\begin{equation}
0<A_2'={1\over U^2}A'({a_{2}^*\over U  })\le 
{1\over U_2^2}A'({a_{2}^*\over U_2})<0.7.
\label{P:4.9}\end{equation}
Thus,
\begin{equation}
{\d^2\ti\F\over\d U\d \a}>0,\quad {\d^2\ti\F\over\d U\d V}<0,
\label{P:4.10}\end{equation}
and using  (\ref{P:4.4}) -(\ref{P:4.10}), we can see immediately
that $\ds{{\d^2\Phi\over\d U \d \a}>0}$. To obtain the second inequality
in (\ref{P:4.2}) we write, using (\ref{P:4.7}) - (\ref{P:4.10}),
$$
0<{-{\d^2\ti\F\over\d U\d V}\over
{\d^2\ti\F\over\d  V^2}}<{1+\de a_{1}^*A_1'\over (1-\de)(1-U^2A_2')}<
{1+1.25\de U^{-2}\over (1-\de)(1-U^2A_2')}\le 1.5,
$$
where we have used also that $U^2A_{1,2}'<1$, bounds (\ref{bounda})
for  $a_{1,2}^*$ and $0.25<U<0.41$. 
Then,
 $$\begin{array}{c}\ds{
{\d^2\Phi\over \d U \d q}={-{\d^2\ti\F\over\d U\d V}\over
{\d^2\ti\F\over\d  V^2}}(\de U A_1'-(1-\de)U A_2')}\\
\ds{
+{\de\over U}A_1-
{(1-\de)\over U}A_2+{\de\over U}a_{1}^*A_1'-
{(1-\de)\over U}a_{2}^*A_2'}\\
\ds{
>{(1-\de)\over U}[A_2'(-a_2^*-1.5U^2)-A_2]> {(1-\de)\over U}[0.5A_2'-A_2]}\\
\ds{
=0.5{(1-\de)A_2\over U}[(A_2-{a_{2}^*\over U^2}+{V\over U}-2]>0}.
\end{array}$$
Thus, we have finished the proof of the first line of (\ref{P:4.1}).

To prove the second line we use the simple statement
\begin{remark}\label{rem:6}
If $f_0(x)=\min_yg(x,y)$ and $\ds{{\d^2g\over \d x^2}\le 0}$, then also
$\ds{{\d^2f_0\over \d x^2}\le 0}$.
\end{remark}
This statement can be easily proved on the basis
of the characteristic property of the concave functions $\ds{{f(x_1)+
f(x_2)\over 2}\le f({x_1+x_2\over 2})}$.

\smallskip

Then on the basis of the second line of (\ref{P:4.7}) we get automatically that
$\ds{{\d^2\Phi\over \d \a^2}\le 0}$.
Therefore, using (\ref{T3.18}) and (\ref{P:4.2}), we get
$$
0<{\d\Phi\over \d \a}(U_1,0,\a_2)<{\d\Phi\over \d \a}(U,0,\a_2)<
{\d\Phi\over \d \a}(U,0,\a).
$$
And so
\begin{equation}
\Phi(U,0,\a)<\Phi(U,0,\a_2)\le\Phi(U(0,\a_2),0,\a_2)<0.
\label{P:4.11}\end{equation}
Now, observing that $\ds{{\d^2\Phi\over \d q^2}\le 0}$
(see Remark \ref{rem:6}),
we conclude that the second line of (\ref{P:4.1}) follows from
(\ref{P:4.11}), if we prove also that for $U\in I_U,\,\a\in I_\a$,
\begin{equation}
{\d\Phi\over \d q}(U,0,\a)<0.
\label{P:4.12}\end{equation}
But since we have proved above that $\ds{{\d^2\Phi\over \d q\d U}>0}$ it is
enough to prove (\ref{P:4.12}) only for $U=U_2$.

The second inequality in   (\ref{T3.18}) implies that
\begin{equation}
{A_2(U_2,0,\a_2)\over A_1(U_2,0,\a_2)}<{\de\over 1-\de}.
\label{P:4.13}\end{equation}
But
$$\begin{array}{c}\ds{
{d\over d\a}{A_2\over A_1}=({1\over U}-V'_\a){A_2\over A_1}({A_2'\over A_2}-
{A_1'\over A_1})}\\
\ds{
=({1\over U}-V'_\a){A_2\over A_1}((A(x_2)-x_2)-(A(x_1)-x_1)),}
\end{array}$$
where $\ds{x_{1,2}={a_{1,2}^*\over U}-V(U,q,\a)}$. Since  $A(x)-x$ is a decreasing
 function (see (\ref{ineqA})) and $U^{-1}-V'_\a>0$ (see (\ref{P:4.5}) and 
(\ref{P:4.7})), we have got that
$$
{A_2(U_2,0,\a)\over A_1(U_2,0,\a)}<{A_2(U_2,0,\a_2)\over A_1(U_2,0,\a_2)}<
{\de\over 1-\de}\,\,\Leftrightarrow\,\,
{\d\Phi\over \d q}(U_2,0,\a)<0.
$$
Thus we have proved the first statement of Proposition \ref{pro:4}. 

Now we are left to prove that inequalities (\ref{T3.18}) and (\ref{T3.17})
implies (\ref{T:3.cond}). To this end it is enough to check that for
$\de\le k_c\a^2$ and $U>\sqrt\a$, $D(U,V(U))\ge 0$, because in this
case we have that ${\cal F}^{(D)}(U,V(U))={\cal F}_0(U,V(U))$ ($U>\sqrt\a$) 
and so
$$
\max_{U\ge\sqrt\a}{\cal F}^{(D)}(U,V(U);q,-q,\a,\de)+C^*(\delta)-
{\a\over 2}\log\a+{\a\over 2}=\max_{U\ge\sqrt\a}\Phi(U,q,\a,\de).
$$
For $U>0.5$ evidently $D(U,V(U);\de)>0$.
 For $0.5>U>\sqrt\a$ we have
 $$\begin{array}{c}
D(U,V(U);\de)>D(\sqrt\a,V(\sqrt\a);\de)\\
\ge D(\sqrt\a,V(\sqrt\a);k_c\a^2)\ge 
D(\sqrt{\a_c},V(\sqrt{\a_c});\de_c).
\end{array}$$
  So, checking numerically that
$D(\sqrt\a_c,V(\sqrt\a_c );\de_c)>0$ we finish the proof of Proposition \ref{pro:4}.
 
\bigskip

\no{\bf Acknowledgements.} This work has been done with the support of
Royal Society and with the help of a scientific agreement between
the Institute for Low Temperature Physics Ukr. Ac. Sci and
the University ``La Sapienza'' of Rome.

\end{document}